\documentclass[aps,12pt,amsmath,amssymb]{revtex4}

\usepackage{hyperref}
\usepackage{graphicx}
\usepackage{dcolumn}   
\usepackage{bm}        
\usepackage{color}
\usepackage{flafter}  

\newcommand{\fj}{f^{(j)}}

\usepackage{amsmath}
\usepackage{amssymb}
\usepackage{makeidx}
\usepackage{graphicx}
\usepackage{color}

\newcommand{\wrt}{w.r.t.\ }

\newcommand{\hide}[1]{}

\newcommand{\up}[1]{ ^{(#1)}}
\newcommand{\inv}[1]{ ^{-#1}}

\newcommand{\myvec}[1]{\vec{#1}}
\renewcommand{\vr}{{\myvec{r}}}

\newcommand{\vA}{{\myvec{A}}}

\newcommand{\vC}{{\myvec{C}}}

\newcommand{\vPsi}{{\myvec{\Psi}}}

\newcommand{\mymat}[1]{{\widehat{#1}}}

\newcommand{\mH}{\mymat{H}}

\newcommand{\mF}{\mymat{F}}

\newcommand{\mD}{\mymat{D}}

\newcommand{\mS}{\mymat{S}}
\newcommand{\mT}{\mymat{T}}

\newcommand{\RR}{\mathbb{R}}
 
\newcommand{\one}{\mathbf{1}} 

\renewcommand{\r}{\rangle}
\renewcommand{\l}{\langle}

\newcommand{\ddt}{\frac{d}{dt}}

\newcommand{\ddx}{\partial_x}
\newcommand{\ddxx}{\partial_x^2}
\newcommand{\ddyy}{\partial_y^2}
\newcommand{\ddy}{\partial_y}

\newcommand{\om}{\omega}

\newcommand{\ep}{\epsilon}

\newcommand{\la}{\lambda}
\newcommand{\La}{\Lambda}

\newcommand{\ga}{\gamma}

\newcommand{\De}{\Delta}
\renewcommand{\th}{\theta}

\newcommand{\lcase}{\left\{\begin{array}{ll}}
\newcommand{\rcase}{\end{array}\right.}
\renewcommand{\bar}{\begin{array}{ll}}
\newcommand{\ear}{\end{array}}
\newcommand{\bma}{\begin{pmatrix}}
\newcommand{\ema}{\end{pmatrix}}
\newcommand{\beq}{\begin{equation}}
\newcommand{\eeq}{\end{equation}}
\newcommand{\bel}[1]{\begin{equation}\label{eq:#1}}
\newcommand{\eel}{\end{equation}}
\newcommand{\bea}{\begin{eqnarray}}
\newcommand{\eea}{\end{eqnarray}}
\newcommand{\beaNN}{\begin{eqnarray*}}
\newcommand{\eeaNN}{\end{eqnarray*}}

\newcounter{lecture}

\newcommand{\vEf}{\vec{\mathcal{E}}}

\newcommand{\vna}{\vec{\nabla}}

\begin{document}

\title{
Perfect absorption in Schr\"odinger-like problems
using non-equidistant complex grids
}

\author{Markus Weinm\"uller, Michael Weinm\"uller, Jonathan Rohland, and Armin Scrinzi} \email{armin.scrinzi@lmu.de}
 \affiliation{Physics Department,
  Ludwig Maximilians Universit\"at, D-80333 Munich, Germany}


\date{\today}

\begin{abstract}
Two non-equidistant grid implementations of infinite range exterior complex scaling
are introduced that allow for perfect absorption in the time dependent Schr\"odinger equation. 
Finite element discrete variables grid discretizations provide as efficient absorption as the corresponding finite elements basis set
discretizations.
This finding is at variance with results reported in literature [L. Tao {\it et al.}, Phys. Rev. A{\bf 48}, 063419 (2009)].
For finite differences, a new class of generalized $Q$-point schemes for non-equidistant grids is derived.
Convergence of absorption is exponential $\sim \Delta x^{Q-1}$ and numerically robust.  Local relative errors $\lesssim10\inv{9}$
are achieved in a standard problem of strong-field ionization. 
\end{abstract}

\maketitle
\section{Introduction}
In the numerical solution of partial differential equations (PDEs) for physical problems 
that involve scattering or dissociation one usually tries to restrict the 
actual computation to a small inner domain and dispose of the parts of the solution
that propagate to large distances. 
The art of achieving this without corrupting the solution in the domain of interest 
is called to ``impose absorbing boundary conditions''. Even without considering questions arising from
discretization it is difficult to lay out a method that would provide perfect absorption 
in the mathematical sense. By perfect absorption we mean a transformation of the original 
PDE defined through an operator $D$ to a new one $D_a$ such that their respective solutions 
$\Psi$ and $\Psi_a$ agree in the inner domain and that $\Psi_a$ is exponentially damped 
in the outer domain. For reasons of computational efficiency we usually require $D_a$ to be
local, i.e. composed of differential and multiplication operators, assuming that $D$ is local. Without this
requirement one can often resort to spectral decomposition of $D$ and apply the desired
manipulations to each spectral component separately to obtain $D_a$. However, this in general 
involves non-local operations with a large penalty in computational efficiency. 

Three local absorption methods for Schr\"odinger-like equations are particularly wide-spread
in the physics community: complex absorbing potentials (CAPs), mask function absorption (MFA),
and exterior complex scaling (ECS). In CAP one adds a complex potential, symbolically written as $D_a=D+V_a$,
that is zero in the inner domain and causes exponential damping of the solution outside. 
The transition from the inner to the outer domain is smoothed to suppress reflections from the
transition boundary. Such potentials are easy to implement and can be rather efficient, in particular
if the spectral range that needs to be absorbed is limited. We restrict the definition of CAP to 
proper potentials, i.e. multiplicative operators that do not involve differentiations. CAPs of this 
kind are never perfect absorbers as defined above.

MFA is arguably the most straight-forward idea: at each time step 
one multiplies the solution by some mask function that is smaller than 1 in the outer domain. 
In the limit of small reduction at frequent intervals this clearly approximates
exponential damping in time. Also, as the solution propagates further into the absorbing
layer, this translates to exponential damping in space. As in CAPs, the mask functions usually 
depart smoothly from the value of 1 at the boundary of the inner domain to smaller values, often
to zero, at some finite distance. In many situations MFA can be understood as a 
discrete version of a purely imaginary CAP $V_a$ by defining the mask function as 
\beq
M_a(x) = \exp[-i\Delta t V_a(x)],
\eeq
where $\Delta t$ is the time step. In such cases one finds similar numerical behavior for both 
methods and the choice between MFA and CAP is only a matter of computational convenience. 

ECS is somewhat set apart from the first two methods in that it systematically derives $D_a$ from 
$D$ by analytic continuation, trying to maintain the desired properties. If one succeeds,
one obtains a perfect absorber in the mathematical sense \cite{mccurdy91:complex_scaling_tdse,scrinzi10:irecs,scrinzi14:ecs-pml}. 
This can be proven for stationary Schr\"odinger operators with free or Coulomb-like 
asymptotics and it has been demonstrated numerically for an important class of linear 
Schr\"odinger operators involving time dependent interactions \cite{scrinzi10:irecs}. The method will be 
discussed in more detail below.

In spite of being ``perfect'' and, as we will demonstrate below, also highly efficient,
ECS has remained the least popular of the three methods, although we may be observing
a recent surge in its application \cite{telnov13:ecs,dujardin14:ecs,deGiovanni15:absorbers,miller14:ecs}.  
The rare use may be related to the fact the ECS requires more care in the implementation 
than CAP and MFA. In fact, to the present date, the most efficient implementations have been
by a particular choice of high order finite elements (FEMs), named ``infinite range ECS'' (irECS) \cite{scrinzi10:irecs}, 
and other local basis sets such as B-splines \cite{saenz:private}. Such methods involve somewhat 
higher programming complexity than grid methods and, more importantly, pose greater 
challenges for scalable parallelization.

When ECS is used in grid methods one usually introduces a smooth transition
from the inner to the absorbing domain, sometimes abbreviated as smooth ECS (sECS, \cite{rom90:smoothECS}). 
Reports about the efficiency of such an approaches appear to be mixed, usually poorer 
than in the FEM implementation, and certainly the results do not deserve the attribute
``perfect''. We will state and demonstrate below what one should expect from 
a perfect absorber in computational practice. The lack of perfect absorbers for grid methods is particularly 
deplorable, as grids are usually easily programmed and as they can also easily be applied to 
non-linear problems.

In this paper we overcome the limitation of irECS to the FEM method 
and introduce grid-implementations that are comparable in absorption efficiency 
with the original irECS implementation, while maintaining the 
scalability of grid methods. We discuss two independent classes of grids: the FEM-discrete
variable representations (FEM-DVR) and finite difference (FD) schemes. The FEM-DVR is a straight 
forward extension of FEM and contrary to earlier reports in literature, it is fully
compatible with ECS and irECS.

For FD we introduce a new approach to non-equidistant grids which maintains the full 
consistency order of FD also for abrupt changes in the grid spacing. 
ECS on grids can be understood as a (smooth or abrupt) transition to a complex spaced
grid in the absorbing region. The irECS scheme correspond to using an exponential complex 
grids for absorption.

Apart from plausibly deriving the schemes, we demonstrate all claims numerically on 
a representative model of laser atom interaction. The computer code and example inputs have been 
made publicly available at \cite{tRecXweb}. 
In particular we will show that the consistency 
order of the grid schemes is as expected $\sim \Delta x^{Q-1}$, where $Q$ is the number of points
involved in computing the derivative at a given point, and that $Q$ can be increased to 
approach machine precision accurate results without any notable numerical instabilities.  
We will show that one can construct such schemes even when the solutions are discontinuous 
and that smooth ECS bears no advantage over an abrupt transition from inner to absorbing domain. 
 
The paper is organized as follows: a brief summary of ECS is given and the FEM irECS implementation
is laid out. Next we show that FEM-DVR is obtained from FEM simply by admitting minor integration 
errors. We will show that these errors do not compromise the absorption properties of irECS.
The second part of the paper is devoted to the new FD schemes applicable for non-equidistant
grids in general and for irECS absorption in particular.

\section{Exterior complex scaling}
\label{sec:scaling}
Exterior complex scaling
is a transformation from the original norm-conserving time dependent Schr\"odinger equation
with a time dependent Hamiltonian $H(t)$ and solution $\Psi(x,t)$ to an equation of the form
\beq
i\ddt \Psi_a(x,t)=H_a(t)\Psi_a(x,t)
\eeq
where solutions $\Psi_a$ become exponentially damped outside some finite region, while inside
the finite region the solution remains strictly unchanged: $\Psi(x,t)\equiv\Psi_a(x,t)$ for 
$|x|<R_0$ and for all $t$. Although, to our knowledge, rigorous
mathematical proof for this fact is still lacking, convincing numerical evidence for the important
class of time dependent Hamiltonians with minimal coupling to a dipole field has been provided \cite{scrinzi10:irecs}.
Apart from these fundamental mathematical properties, in practical application it is important 
that machine precision accuracy can be achieved with comparatively little numerical effort.
The particular discretization scheme that provides this efficiency was dubbed ``infinite range 
exterior complex scaling'' (irECS). In Ref.~\cite{scrinzi10:irecs} it is shown that,
comparing with a popular class of complex absorbing potentials (CAPs), the irECS scheme
provides up to 10 orders of magnitude better accuracy with only a fraction of the absorbing boundary size. 
This original formulation of irECS was given in terms of a finite element discretization.
To our knowledge (and surprise) irECS has not been implemented by other
practitioners, although its good performance appears to have lead to a 
re-assessment of ECS methods for absorption and encouraging results were reported  
\cite{telnov13:ecs,dujardin14:ecs,miller14:ecs,deGiovanni15:absorbers}.

In this section we give a brief formulation of ECS that will allow us to formulate the essential 
requirements for a numerical implementation.

\subsection{Real scaling}
Exterior complex scaling derives from a unitary scaling transformation $U_a$ from the original coordinates $x$
to new coordinates $y$ which is defined as 
\beq\label{eq:scalingU}
(U_a\Psi)(y)=\la(y)^{1/2}\Psi(\La(y)),\quad \La(y):= \int_{-\infty}^y \la(s)ds
\eeq
with a real scaling function
\beq
0<\la(y) = \lcase 1&\text{for } |y|<R_0\\ a\, g(y) & \text{for } |y|\geq R_0\rcase.
\eeq 
The transformation is unitary for any positive $g(y)>0$ and any positive $a$. 
One sees that the transformation leaves $(U_a\Psi)(y)=\Psi(y)$ invariant in the inner domain $y<R_0$ and that
it stretches or shrinks the coordinates for $a g(y)\gtrless 1$. 
Note that here $\la(y)$ only is required to be positive, but no continuity assumptions are made.  

Switching from the Schr\"odinger to the Heisenberg picture and considering $U_a$ as a transformation 
of operators rather than wave functions, we can define a scaled Hamiltonian operator
\beq
H_a = U_a H U^*_a.
\eeq
Clearly, as a unitary transformation $U_a$ leaves all physical properties of the equation invariant. One can 
think of the transformation as the use of locally adapted units of length. 

Some caution has to be exercised, when  $\la(y)$ is non-differentiable or discontinuous. 
Obviously, starting from a differentiable $\Psi$ the corresponding $\Psi_a$ 
will become non-differentiable or discontinuous to the same extent as $\la(y)$ is non-differentiable or
discontinuous.
As a result,
we cannot apply the standard differential operators $\Delta$ or $\vna$ to $\Psi_a$.
Of course, by construction we can apply the transformed $\Delta_a = U_a\Delta U_a^*$ 
and $\vna_a= U_a\Delta U_a^*$ to it. Conversely, the transformed $-\Delta_a$ 
and $\nabla_a$ cannot be applied to the usual differentiable functions $\Psi$, but only 
to functions obtained from differentiable functions by the transformation $\Psi_a=U_a\Psi$.
This simple observation will be the key to constructing numerically efficient 
discretization schemes for the scaled equations and also to write all transformed 
discretization operators in the simplest possible form. 

A short calculation shows that the transformed first and 2nd 
derivatives have the form 
\bea
i\nabla_a &=& i\la(y)\,\inv{1/2}\nabla\la(y)\inv{1/2} \left[=\frac{i}{a} g(y)\inv{1/2}\nabla g(y)\inv{1/2}\right]_{|y|>R_0}\\
\De_a &=& \la(y)\inv{1/2}\nabla\la(y)\inv1\nabla\la(y)\inv{1/2} \left[= \frac{1}{a^2} g(y)\inv{1/2}\nabla g(y)\inv1\nabla g(y)\inv{1/2}\right]_{|y|>R_0},
\label{eq:ddxxTrans}
\eea
here written in a manifestly Hermitian form. Potentials simply transform by substituting for the argument:
\beq
V_a(y) = V(\La(y)).
\eeq

\subsection{Complex scaling}

Complex scaling consists in admitting complex $a$ with $\Im(a)>0$. To see how this leads to exponential damping
of the solution, one can consider the transformation of outgoing waves $k>0$ at $x\to \infty$ (assuming
for simplicity $g(y)\equiv 1$)
\beq
\Psi(x)\sim e^{ikx}\to \Psi_a(y) \sim a^{1/2} e^{iky\Re(a) - ky\Im(a)}.
\eeq
Ingoing waves $k<0$ would be exponentially growing and are excluded, if we admit only
square-integrable functions in our calculations. This is the case if we calculate on a finite
simulation box $[-L,L]$ with Dirichlet boundary conditions at $\pm L$. 

For defining $V_a$ at complex values of $a$, there must be an analytic continuation
of $V(x)$ to complex arguments $V_a(y)=V(\La(y))$ in the outer domain $x>R_0$. 
For being useful in scattering situations, the analytic continuation must maintain the asymptotic 
properties of the potential, such as whether it admits continuous or only strictly 
bound states. This is not guaranteed: for example, for $\arg(a)>\pi/4$,
complex scaling turns the harmonic potential from confining $\propto x^2$ to repulsive $\propto -y^2$. 
No such accident happens in typical systems showing break-up or scattering with Coulomb or free asymptotics.
A much more profound discussion of the mathematical conditions for complex scaling can be found in \cite{reed_simon82:complex_scaling}.

Apart from the exponential damping of the solutions, the second important property for application of ECS
in time dependent problems is stability of the time evolution: the complex scaled
Hamiltonian $H_a(t)$ must not have any eigenvalues in the upper half of the complex plane. If such eigenvalues appear, they
will invariably amplify any numerical noise as the solution proceeds forward in time 
and as a result the complex scaled solution $\Psi_a(t)$ will diverge. 
Luckily, stability has been shown for a large class of Schr\"odinger-type equations. 
This includes time dependent Hamiltonians with velocity gauge coupling $i\vna\cdot\vA(t)$ to a time dependent
dipole vector potential $\vA(t)$. 
It is interesting to note that the length gauge formulation of the same physical problem with the 
coupling $\vr\cdot\vEf(t)$ in instable under ECS \cite{mccurdy91:complex_scaling_tdse}. This may be surprising, considering that the two
forms are related by a unitary gauge transformation. However, as the gauge transformation is 
space-dependent, its complex-scaled counterpart is not unitary and 
changes the spectral properties of the Hamiltonian. 

For a more detailed discussion as to why the solution remains invariant in the inner domain and under which 
conditions time-propagation of the complex scaled system is stable, we refer to \cite{scrinzi14:ecs-pml} and 
references therein.

\section{The FEM-DVR method for irECS}
Here we lay out how ECS is implemented for FEM-DVR methods. The FEM-DVR approach 
was introduced in \cite{manolopoulos88:femdvr}. Mathematically it differs from a standard finite 
element method only by the admission of a small quadrature error. Therefore we first formulate 
the standard finite element method in a suitable way. We limit the discussion to the one-dimensional case. 
Extensions to higher dimensions are straight forward
and the problems arising are not specific to the individual methods. 

\subsection{A formulation of the finite element method}

In a one-dimensional finite element method one approximates some solution $\Psi(x)$, $x\in\RR$ 
piecewise on $N$ intervals, the finite elements $[x_{n-1},x_n], n=1,\ldots N$.
With local basis functions $f\up{n}_k(x),\ldots k=1,\ldots,Q_n$ that are zero outside the $[x_{n-1},x_n]$
one makes the ansatz
\beq\label{eq:femansatz}
\Psi(x)=\sum_{n=1}^N \sum_{k=1}^{Q_n} f\up{n}_k(x) c_{nk}=:\vec{|F\r}\cdot\vC.
\eeq
Note that interval boundaries $x_n$, number $Q_n$, and type of 
functions $f\up{n}_k$ can be chosen without any particular constraints, except for 
the usual requirements of differentiability, linear independency, and completeness in the limit $Q_n\to\infty$. 
If the exact solution $\Psi$ is smoothly differentiable,
polynomials are a standard choice for $f\up{n}_k$. However, at specific locations or at large $|x|$ 
other choices can bear great numerical advantage, as will be discussed below.
Equations of motion for the $\vC$ are derived by a Galerkin criterion 
(in physics usually called the Dirac-Frenkel variational principle) with the result 
\beq\label{eq:discreteTDSE}
i\ddt \mS \vC(t) = \mH(t) \vC(t),
\eeq
where the Hamiltonian $\mH(c)$ and overlap $\mS$ matrices are composed of the piece-wise matrices
\bea
\mH\up{n}_{kl}&=&\int_{x_{n-1}}^{x_n} dx f\up{n}_k(x) H(t) f\up{n}_l(x)\\
\mS\up{n}_{kl}&=&\int_{x_{n-1}}^{x_n} dx f\up{n}_k(x) f\up{n}_l(x).
\eea
Here and in the following we assume that the $f\up{n}_k$ are real-valued. Complex functions used in practice,
such as spherical harmonics, can be usually obtained from purely real functions by simple linear transformations.
Clearly, we assume that $H(t)$ is local, i.e. matrix elements of functions from 
different elements $n\neq n'$ are $\equiv0$. Note that all basis sets that are related by a $Q_n\times Q_n$ similarity
transformation $\tilde{f}\up{n}_{k'}=\sum_k \mT\up{n}_{k'k}f\up{n}_k$ are mathematically equivalent. 
These $Q_n^2$ free parameters can be used to bring the $f\up{n}_k$ to a computationally
and numerically convenient form.

So far, the ansatz (\ref{eq:femansatz}) admits $\Psi(x)$ that are not
differentiable or even discontinuous at the element boundaries $x_n$. 
It is well known that for a correct definition of the discretization of the differential operators $i\ddx$ and $\ddxx$ 
it is sufficient to ensure that the $\Psi(x)$ are {\em continuous} at $x_n$, 
if one secures that all operators are implemented in a manifestly symmetric form.
The correct symmetric form is typically obtained by a formal partial integration where boundary terms are dropped 
(see, e.g., \cite{scrinzi:jcp1993} for a more detailed derivation). For example
\beq\label{eq:symm1}
-\l f\up{n}_k|\ddxx f\up{n}_l\r\to \l \ddx f\up{n}_k|\ddx f\up{n}_l\r.
\eeq
Note that explicit (anti-)symmetrization must also be observed for operators involving first derivatives, e.g.
\beq\label{eq:symm2}
\l f\up{n}_k|(g\ddx+\frac{g'}{2})f\up{n}_l\r\to \frac12 \left(\l f\up{n}_k|g\ddx f\up{n}_l\r-\l g\ddx f\up{n}_k| f\up{n}_l\r\right).
\eeq

Continuity can be most conveniently realized by applying a similarity transformation $\mT\up{n}$ on each element such
that only the first and the last function on the element are non-zero at the interval boundaries and fixing these boundary
values to 1:
\beq\label{eq:fnbasis}
f\up{n}_1(x_{n-1})=f\up{n}_{Q_n}(x_{n})=1, \text{ else }f\up{n}_k(x_{n-1})=f\up{n}_{k}(x_{n})=0.
\eeq
Implementation of these $2Q_n$ conditions fixes only $2Q_n$ out of the $Q_n^2$ free parameters in $\mT\up{n}$.
The remaining freedom can be used for further transforming the basis set. 

With such functions continuity can be imposed by simply setting equal the coefficients $c_{nk}$ corresponding to the
left and right functions at each element boundary $x_n$:
\beq
c_{n,Q}\equiv c_{n+1,1}\quad\forall n.
\eeq
In the full matrices $\mS$ and $\mH(t)$ the corresponding rows and columns will be merged. 
One readily sees this amounts to adding $\mS\up{n}$ and $\mH\up{n}$ into $\mS$ and $\mH(t)$
such that the lower right corners of the $n$'th submatrix overlaps with the upper left corner of the $n+1$st matrix 
(for an illustration, see, e.g.\cite{scrinzi10:irecs}).

In general, the $\mS\up{n}$ will be full. One can use the remaining freedom in $\mT\up{n}$ to bring the 
matrices $\mS\up{n}$ to nearly diagonal form where only two non-zero off-diagonal elements remain and there are all 1's 
on the diagonal except for the first and the last diagonal entry. Complete diagonalization of $\mS$ 
is inherently impossible without destroying the locality of the FEM basis.

The non-diagonal form of $\mS$ is the primary technical difference between grid methods and FEM. It is a significant drawback, 
in particular, when operating on parallel machines, where either iterative methods must be employed or all-to-all communication is required.
This is not a problem of operations count: applying the inverse in its near-diagonal form 
with only two off-diagonal elements for each of the $N$ elements can be reduced to solving a tri-diagonal linear
system of size $N-1$. However, solving the tridiagonal system connects all elements to each other. In a parallel 
code where the elements are distributed over compute nodes
this ensues costly all-to-all type of communication and may require complex coding, especially in higher dimensions.

\subsection{A formulation of the FEM-DVR method}
In FEM-DVR one reduces the overlap matrix to $\mS=\one$ by admitting a small quadrature error in the computation
of matrix elements. We introduce the FEM-DVR discretization based on the approach above. We choose our functions
in the form
\beq
f\up{n}_k(x) = p\up{n}_k(x) v\up{n}(x),
\eeq 
with polynomials of maximal degree $Q_n-1$ for $p_k(x)$ and a weight function $v\up{n}(x)$.
For such functions there is a $Q_n$-point Lobatto quadrature rule
\beq
\int_{x_{n-1}}^{x_n} ds f\up{n}_k(s)f\up{n}_l(s)  \approx \sum_{i=1}^{Q_n} w_i p\up{n}_k(s_i) p\up{n}_l(s_i),
\eeq
where the quadrature points include the interval boundary values $x_{n-1}=s_1<s_1<\ldots<s_{Q_n}=x_n$.
We can construct our basis functions $f\up{n}_k$ using the Lagrange polynomials for the Lobatto quadrature points $s_i$
as
\beq
f\up{n}_k(x) =\frac{v\up{n}(x)}{v\up{n}(s_k)}\prod_{i\neq k} \frac{x-s_i}{s_k-s_i}, 
\eeq
which have the properties  (\ref{eq:discreteTDSE}). If instead of exact integration one contents oneself with
(approximate) Lobatto quadrature one finds a diagonal overlap matrix:
\beq
\mS\up{n}_{kl} = \l f_k\up{n}| f_l\up{n}\r\approx \sum_{i=1}^{Q_n} w_i p_k(s_i) p_l(s_i) = w_k \delta_{kl}. 
\eeq
In fact, exact integration is only missed by one polynomial degree, as Lobatto quadrature is exact
up to degree $2Q_n-3$, while our $p_k\up{n}$ are degree $Q_n-1$. The two degrees lower accuracy 
of Lobatto compared to standard Gauss quadrature is the penalty for fixing $s_1$ and $s_{Q_n}$ to coincide with 
the interval boundaries.

A further advantage of FEM-DVR over FEM is that Lobatto quadrature is applied for all multiplicative operators, not only
the overlap. By that all multiplication operators
are strictly diagonal and allow highly efficient application. The advantage is mostly played out in higher dimensions, 
where the exact basis set representation of a general potential would be a full matrix. 
Derivative operators are full in FEM-DVR, but usually they come in the form of a short sum of tensor products, 
which again can be implemented efficiently.

\subsection{ECS and irECS for a FEM-DVR grid}

The favorable absorption properties of ECS in general and of the particular implementation 
by irECS were first reported in Ref.~\cite{scrinzi10:irecs} and have since be used to solve several
challenging problems in the strong laser-matter interactions 
\cite{hofmann14:elliptic,majety15:hacc,majety15:exchange,zielinski14:fanoArXiv,torlina15:attoclock}.
All these calculations were performed using a FEM basis.

In fact, in Ref.~\cite{tao09:complex_scaling} severe instabilities were reported for ECS absorption using FEM-DVR discretization
for a simple test-problem where the irECS showed perfect absorption. In Ref.~\cite{scrinzi10:irecs} we speculated that the
the approximate quadratures inherent to FEM-DVR were to blame. This is a plausible possibility,
as analyticity plays a crucial role for ECS and small integration errors
by using Lobatto quadrature instead of evaluating integrals exactly might destroy perfect absorption.
Now we show that this speculation was incorrect, that FEM-DVR gives numerical results of the same quality as FEM, 
and that the problems encountered in \cite{tao09:complex_scaling} must have had a different origin.

All calculations below were performed using the tRecX code, which together with the relevant example inputs
has been made publicly available \cite{tRecXweb}.

\subsubsection{Model system}
In all numerical examples in this paper 
we use as a model Hamiltonian the ``one-dimensional Hydrogen atom'' in a laser field (using atomic units $\hbar=e=m_e=1$)
\beq\label{eq:modelH}
H(t) = -\frac12\ddxx -iA(t)\ddx -\frac{1}{\sqrt{x^2+2}},
\eeq
where $A(t)$ is the laser field's vector potential (in dipole approximation). 
 Remarkably, the ground state energy of the 
system is exactly at -1/2, as in the three-dimensional Hydrogen atom. It has been demonstrated that
the mathematical behavior of absorption in this simple system generalizes to the analogous Schr\"odinger equations for one-
and two-electrons systems in up to 6 spatial dimensions \cite{majety15:hacc,zielinski15:di}.
For all studies below use a vector potential of the form
\beq
A(t) = A_0\cos^2(\frac{\pi t}{2T})\sin(\om t) \text{ for } t\in[- T, T]
\eeq
with $A_0=1.3\,au$, $\om=0.057$, and $T=3\pi/\om$. In more physical terms, the parameters translate into 
a pulse with central wavelength of $800\,nm$, peak intensity $2\times10^{14}W/cm^2$, and a FWHM duration of three optical cycles.
Such a pulse depletes the initial ground state of the system by about 50\%, which is all absorbed at the boundaries.
In this type of strong-field ionization processes emission occurs over a very wide spectral range. At our parameters
outgoing wave-vectors cover the whole range from zero up to $\sim 2\,au$ before amplitudes drop to below physically relevant levels.
This broad range of outgoing wave vectors poses a particular challenge for absorption.
At more narrowly defined ranges of wave vectors absorption can be achieved by a variety of methods by tuning the parameters
to the specific wave vectors. One of the advantage of ECS is that it can be applied over the whole range without the need
to adjust to the particular form of emission.

An important physical parameter of strong field photo-emission is the ``quiver radius'': 
a classical free electron will oscillate with an amplitude $a_0=A_0/\om$ in the laser field. At our parameters 
one computes a quiver radius of $a_0\approx 23$.
This gives a rough measure for the radius up to where one needs to preserve the solution without absorbing and 
it motivates our choice of $R_0\ge 25$. Note that, if so desired, ECS allows choosing arbitrarily
small $R_0$ (including $R_0=0$) such that flux may propagate deeply into the complex scaled 
domain and return to the inner domain without necessarily corrupting the solution 
(see Ref.~\cite{scrinzi10:irecs} for more details). This fact further corroborates that, mathematically speaking, ECS is
a lossless transformation. The loss of information by the exponential damping is purely numerical due to 
the limited accuracy of any finite representation of the solution.

\subsubsection{Implementation of ECS}
We use the simplest scaling function $g(y)\equiv1$, i.e.
\beq\label{eq:lambda}
\la(y) = a\text{ for }|y|>R_0.
\eeq
The scaled Hamiltonian is
\beq
H_a(t) = -\frac12[\ddyy]_a -iA(t)[\ddy]_a-\frac{1}{\sqrt{\La(y)^2+2}},
\eeq
with the scaled derivatives
\beq
[\ddy]_a,[\ddyy]_a = \lcase \ddy,\ddyy &\text{ on } |y|<R_0\\ \frac{1}{a}\ddy, \frac{1}{a^2}\ddyy,& \text{ on }|y|>R_0. 
\rcase
\eeq
The scaled solution will have the form $U_a\Psi$ for some differentiable $\Psi$. In particular, $\Psi_a$ has a discontinuity 
by the factor $a^{1/2}$ when crossing the scaling radius $R_0$. In a finite element scheme it is easy to implement such a discontinuity:
we choose two element boundaries to coincide with the lower and upper boundaries of 
the inner domain $\pm R_0=y_{n_\pm}$. Then all functions on the outer domain are multiplied by $a^{1/2}$
\beq
f_k\up{n}\to a^{1/2}f_k\up{n}\quad\text{ for } y_{n-1}\geq R_0\text{ or } y_n \leq -R_0.
\eeq 
The desired {\em dis}continuity is ensured by equating the coefficients corresponding to the boundaries $y_{n_\pm}$, 
just as continuity is ensured at all other boundaries.
Conditions on the derivatives can be omitted for the same reasons and with the same precaution about using explicitly 
symmetric forms of the operators as discussed above, Eqs.~(\ref{eq:symm1}) and (\ref{eq:symm2}).

In practical implementation, multiplying the function translates into a multiplication of the 
element matrix blocks $\mS\up{n}$ and  $\mH\up{n}$ by $a$:
\beq
\left.
\bar
\l f_k\up{n}| f_l\up{n}\r &\to a\l f_k\up{n}| f_l\up{n}\r\\
\l f_k\up{n}|H(t) |f_l\up{n}\r &\to a\l f_k\up{n}|H_a(t) |f_l\up{n}\r
\ear
\right\}\text{ for } y_n \leq -R_0 \text{ or }  R_0\leq y_{n-1}.
\eeq 
Obviously, the only non-trivial effect of this extra multiplication by $a$ appears at blocks
to either side of $\pm R_0$. Also note that, while the overlap matrix blocks $\mS\up{n}$ remain
unchanged except for the multiplication by $a$, one must use the properly scaled operators 
$H_a(t)$ for evaluating the scaled matrix blocks $\mH_a\up{n}(t)$. 

Complex scaling now means that the {\em operator} is analytically continued \wrt $a$. There is a seeming ambiguity as how to 
deal with complex conjugation of $a^{1/2}f\up{n}_k$ in the scalar products. One might suspect that in fact
there should be $|a|$ appearing as a factor for the matrices rather than $a$. 
Clearly, this would pose a problem as the modulus is not an analytic function and analytic continuation of the operator 
would be doomed. Closer inspection shows that bra functions $\l \psi|$ must be chosen differently from the ket functions $|\psi\r$, 
exactly such that the conjugation of $a^{1/2}$ is undone, see \cite{combes87:resonance,scrinzi10:irecs,scrinzi14:ecs-pml}. 
Thus, it is the $a$ appearing in the {\em operator} that are extended to complex values.

\subsubsection{irECS discretization}
The irECS version of ECS greatly enhances computational efficiency by replacing the Dirichlet conditions at the finite boundaries $\pm L$
with a computation on an infinite interval where exponentially decaying basis functions ensure decay $\to 0$ as $|y|\to\infty$. 
For our two-sided infinity this amounts
to formally choosing $-y_0=y_N=\infty$ and using the weight functions $v(y)=\exp(\pm \ga y)$ at the first and last interval, respectively.
The finite inner domain of the axis is divided into elements of equal size.
We construct the $f\up{n}_k$ as in (\ref{eq:fnbasis}). It has been investigated earlier how
errors of irECS in the FEM implementation behave with order, number of elements, complex scaling angle $\th$, scaling
radius, and exponential factor $\exp(- \ga |y|)$ on the infinite intervals. Summing up those results, 
irECS absorption is highly efficient and, within reasonable limits, quite insensitive to these details of the discretization. 
For a quali- and quantification of this statement we refer to Ref.~\cite{scrinzi10:irecs}.

The irECS idea readily carries over to FEM-DVR, if we use
a Radau quadrature formula for the infinite intervals. Radau quadrature fixes only one quadrature point at the 
finite left or right boundary of the interval. One adjusts the remaining quadrature points and the quadrature weights for 
the specific weight function $|v(y)|^2$ such that with $Q_n$ quadrature points integrals become exact up to 
polynomial degree $Q_n-2$.

For demonstrating that the numerical behavior of FEM-DVR and FEM are equal for all practical purposes, we use a fixed set
of discretization parameters. We choose the complex scaling parameter $a=\exp(i/2)$ and radius $R_0=25$ 
with 10 intervals of size $5\,a.u.$ in the inner domain $[-R_0,R_0]$. For the infinite element basis we
use the $v(y)=\exp(-|y|/2)$, i.e. $\ga=0.5$. We use a uniform order $Q_n\equiv Q$ in the inner domain 
for all elements  and two infinite elements of order $Q_1=Q_{N}=:A$ in the outer domain to either side.
We study the variation of the results with $Q$ and $A$. The same functions $f_k\up{n}$
with Lagrange polynomials at the Lobatto points (finite elements) or Radau points (infinite elements) are used in FEM and FEM-DVR. 
For FEM-DVR, this choice is by definition. For FEM the exact choice of the polynomials is unimportant: results near 
machine precision can be obtained with any set of 
polynomials, if only one avoids ill-conditioning problems as they typically arise in too simplistic choices, such
as monomials. In fact, in all previous calculations we had derived our basis from Legendre (finite) and Laguerre (infinite range)
polynomials. 
For FEM, we compute all matrix elements to machine precision using a recursive algorithm. In the given basis, 
FEM-DVR simply consists in replacing the exact integrals with $Q$-point Lobatto and $A$-point Radau quadratures on
the respective elements.

Throughout this work we assess the accuracy of the solutions by computing the local and  maximal relative errors of the 
probability density at the end of the pulse $\rho(x):=|\Psi(x,T)|^2$:
\beq
\ep(x):=\frac{\rho(x)-\rho_0(x)}{\rho_0(x)},\quad\ep_0:=\max_{x\in[-R_0,R_0]}\ep(x).
\eeq
The reference density  $\rho_0$ is drawn from a large, fully converged calculation.

For time-integration we use the classical Runge-Kutta scheme with step-size control based on the maximal error of the coefficient
vector components $c_{nk}$. This universally applicable method was selected to facilitate comparisons between the methods, 
without any attempt to optimize its performance.

\begin{figure}
\includegraphics[width=0.8\textwidth]{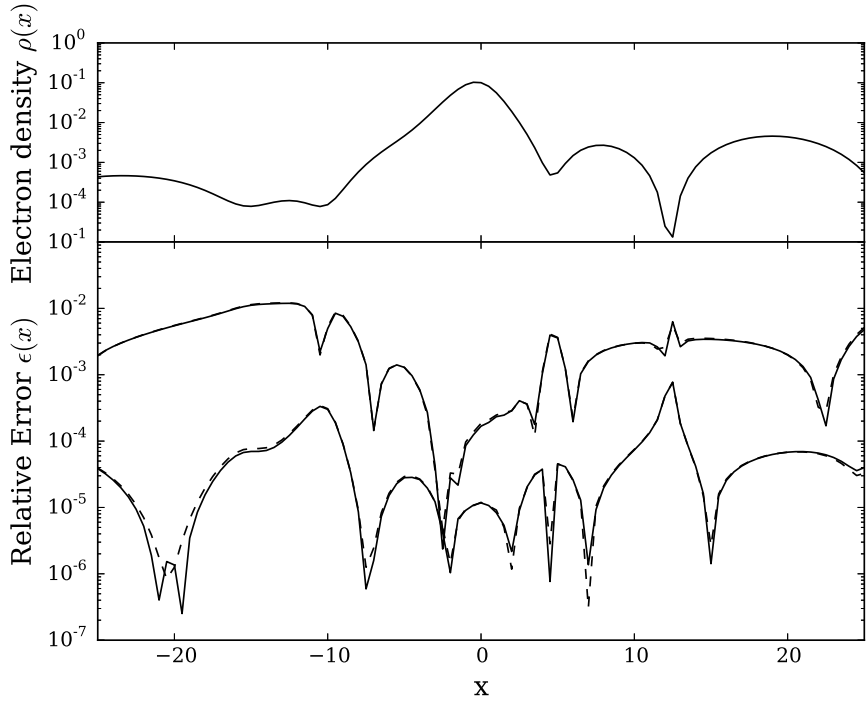}
\caption{\label{fig:densFem}
Density $\rho(x)$ (upper panel) and discretization error $\ep(x)$  
at the end of a 3-cycle laser pulse at 800 nm central wave length and peak intensity $2\times10^{14}$.
The $\ep(x)$ are barely distinguishable,  solid line: FEM, dashed: FEM-DVR. The more accurate result at $\ep(x)\lesssim 10\inv4$ was obtained 
with discretization $(Q,A)=(12,18)$, 186 points, errors $\sim 10\inv2$
are reached with  $(Q,A)=(9,15)$, 150 points. (See text for the definition of $Q,A$ and $\ep(x)$.)
}
\end{figure}

Figure~\ref{fig:densFem} shows $\rho_0(x)$ and the $\ep(x)$ corresponding to two pairs of FEM and FEM-DVR calculations 
with errors $\ep(x)\lesssim 10\inv2$ and  $\ep(x)\lesssim 10\inv4$. 
One observes that FEM and FEM-DVR produce, at equal discretization size, 
equally accurate absorption with no obvious accuracy advantage for either method.

\begin{figure}
\includegraphics[width=\textwidth]{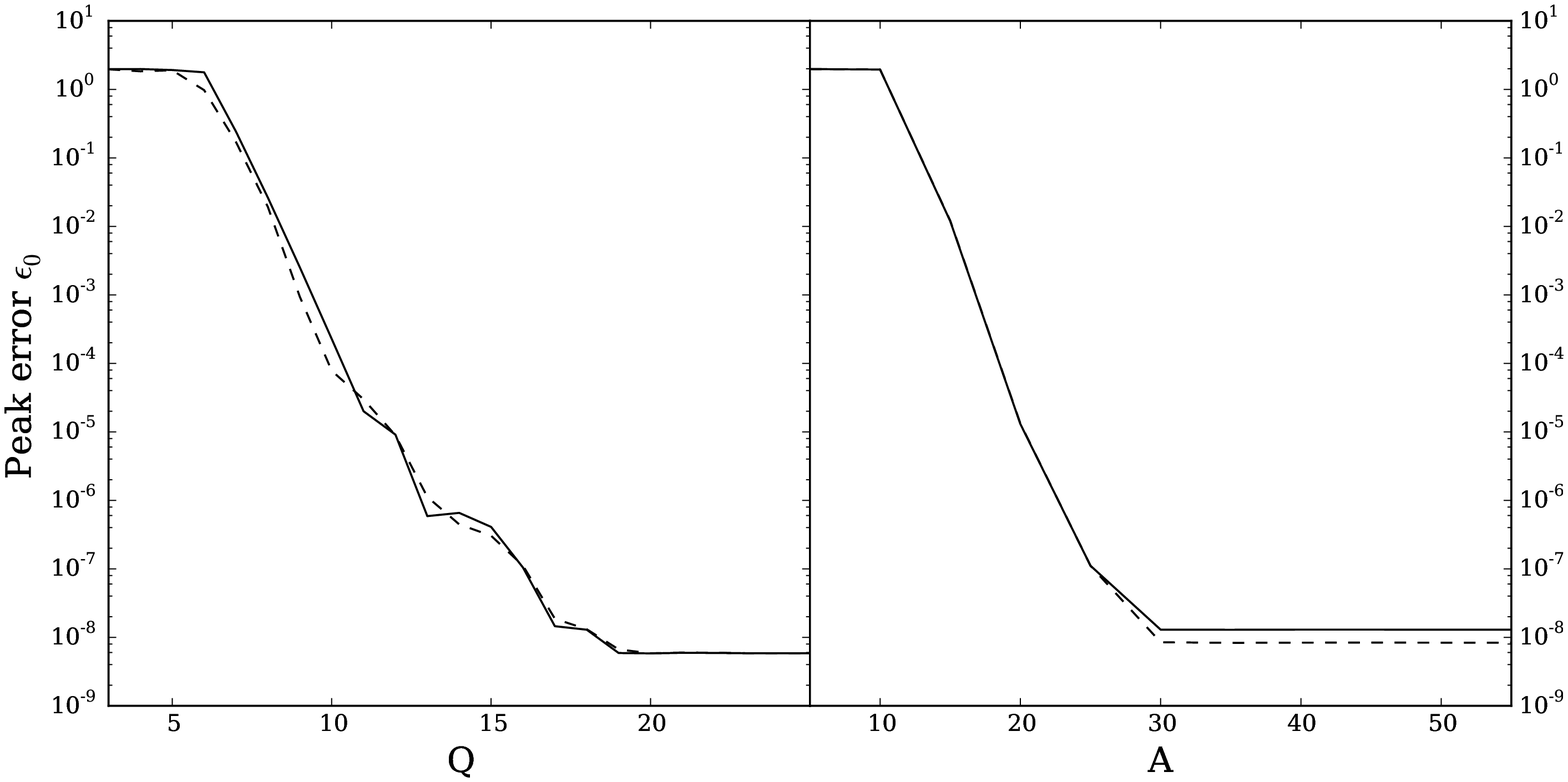}
\caption{\label{fig:dvrconvergence}
Convergence of the peak error $\ep_0$. Left: as function of discretization order $Q$ in the inner domain keeping $A=A_0$ fixed, 
right: as a function of absorption order $A$, keeping $Q\equiv Q_0$ fixed. Solid lines: FEM, dashed FEM-DVR. 
System parameters as in \ref{fig:densFem}. 
}
\end{figure}

Fig.~\ref{fig:dvrconvergence} shows thee discretization errors in the inner domain and complex scaled domain independently. 
Computations are with $Q,A_0$ varying $Q$ for the inner domain and $Q_0,A$ varying $A$ for absorption. The joint parameters $Q_0,A_0$ 
are chosen such that the solution is maximally accurate. As expected, errors drop exponentially with 
$Q$, and also absorption improves exponentially beyond a minimum number of $A\gtrsim 12$.

\subsection{Discussion and conclusions on FEM-DVR}
The simple conclusion of this first part is that FEM-DVR is completely at par with FEM, at least as far as perfect absorption is concerned.
Considering the great simplification for the computation of integrals and, more importantly, the gains in operations count, ease
of implementation, and ease of parallelization it is certainly to be preferred over a full blown FEM method. There is only 
a single point where we see some advantage of computing the integrals exactly: FEM eigenvalue estimates are variational upper bounds,
while FEM-DVR approximations may drop below the true value. The actual errors are, according to our observation, of similar magnitude
in both methods. In practice, the upper bounding property will rarely be of great importance.

\begin{figure}
\includegraphics[width=0.9\textwidth]{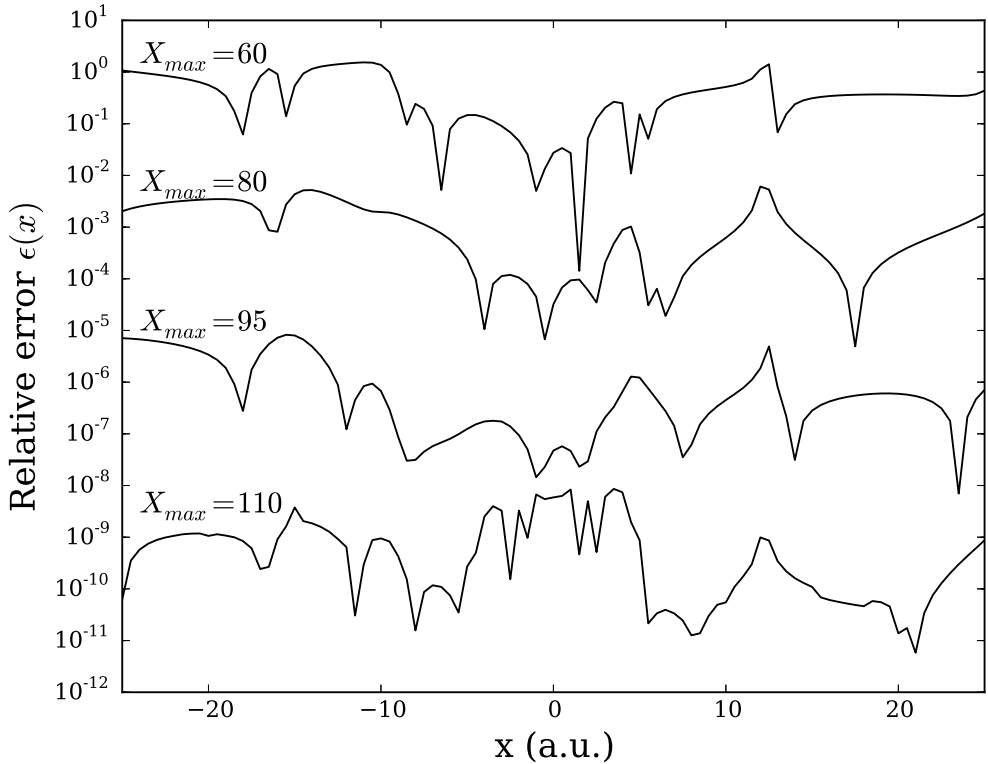}
\caption{\label{fig:ecs}
Local error $\ep(x)$ for absorption in a finite box as the box $[-X_{\rm max},X_{\rm max}]$ is extended by adding elements in the absorbing domain. Size of all 
elements is $x_n-x_{n-1}=5$ and order is fixed at $Q=A=21$.   
}
\end{figure}

Our conclusion is at variance with previous reports on absorption using FEM-DVR. In Ref.~\cite{tao09:complex_scaling} 
the infinite range idea was not used.
However, at some expense as to efficiency, also a standard DVR discretization using only polynomials without exponential damping
produces highly accurate results. This was reported for FEM in \cite{scrinzi10:irecs} and can be reconfirmed for FEM-DVR here. 
Fig.~\ref{fig:ecs} shows
the errors $\ep(x)$ of a discretization where the infinite intervals were replaced by an increasing number of finite elements of the constant 
order $Q_0$. One sees that errors can be very well controlled and no artifacts appear also in the area $|x|\sim R_0$. 
In spite of similar discretization sizes, the calculations in \cite{tao09:complex_scaling} were reported to be instable, 
especially near $|x|=R_0$.

\section{Generalized FD schemes}
In FD schemes there is usually no explicit reference made to an underlying 
basis, but rather one is contented with representing the wavefunction at the grid points. 
In reality also FD uses a hypothesis for evaluating the derivatives: in standard
applications one assumes that in the vicinity of a grid point $x_j$ the solution can be well 
approximated by a linear combination of polynomials $f\up{j}_k(x)$. We first re-derive the 
standard schemes in analogy to the discussion of the FEM method above and then generalize 
the approach for non-differentiable solutions.
 
Standard symmetric FD schemes on equidistant one-dimensional grids $x_j=x_0+jh, j=0,\ldots, J$ are obtained by
assuming that in the vicinity of each point $x_j$ one can write 
\beq\label{eq:hyp0}
\Psi(x)\approx \sum_{k=-P}^{P} \fj_k(x) c_k.
\eeq
The ``interpolation hypothesis'' is that the functions $\fj_k$ are polynomials.   
We treat here only the case of symmetric schemes with an odd number of 
$Q=2P+1$ points and use the notation $\mF\up{j}_{kl}:=\fj_k(x_{j+l})$ with the
index ranges $l,k\in\{-P,\ldots,P\}$.
The coefficients $c_k$ can be obtained from the neighboring function values $\Psi_{j+l}:=\Psi(x_{j+l})$ 
as 
\beq\label{eq:coef0}
c_k =\sum_{l=-P}^P \left[\left(\mF\up{j}\right)\inv1\right]_{kl} \Psi_{j+l}.
\eeq
One readily obtains an approximation to the derivative $(\ddx\Psi)(x_j)=:\Psi'_j$ as
\beq\label{eq:fdstandard}
 \Psi'_j\approx \sum_{l,k} (\ddx\fj_k)(x_j) \left[\left(\mF\up{j}\right)\inv1\right]_{kl} \Psi_{j+l}
=: \sum_{l=-P}^P d\up{j}_l \Psi_{j+l}.
\eeq
Arranging the pointwise finite difference rules $d\up{j}_l$ into a matrix $\mD\up1_{j,j+l}=d\up{j}_l$ 
we find for the finite difference approximation of the first derivative
\beq
\vec{\Psi}'=\mD\up1\vec{\Psi}.
\eeq
The same construction principle can be used for higher derivatives or actually 
any operator composed of derivatives and multiplicative operators.

Eq.~(\ref{eq:fdstandard}) is suitable for the construction of the schemes in numerical practice, 
if only one avoids ill-conditioning of $\mF\up{j}$. Almost any choice  for the $\fj_k$, e.g. orthogonal polynomials 
or Lagrange polynomials at well-separated support points, will suffice.

\subsection{Non-equidistant, discontinuous, and complex scaled schemes}

A standard way of constructing FD schemes for non-equidistant grids is based on the 
very same coordinate scaling discussed in the preceding section. 
Let us assume that the non-equidistant grid is defined by some monotonically
increasing function as
\beq
x_j = \La(y_j) \text{ with } y_j=0,h,2h,\ldots,
\eeq
which transforms the representation on the non-equidistant $x_j$ grid into a representation on the equidistant $y_j$ grid.
For example, for exponential sampling of $x\geq0$ one can choose $\La(y)=\exp(y)-1, j=0,\ldots, N-1$
or $y=\log(1+|x|)$. For deriving the necessary transformation of the operators and for constructing
norm-conserving schemes, it is useful to adhere to the notation of section \ref{sec:scaling}
and consider the coordinate transformation as a unitary transformation $\Psi\to \Psi_a=U_a\Psi$, 
Eq.~(\ref{eq:scalingU}). 

The key to suitable FD schemes in the transformed coordinates is to realize that the 
interpolation hypothesis for the transformed solution in general must differ from the original one.
Suppose a set of functions $\fj_k$ is in some sense an optimal interpolation hypothesis for the unstretched 
$\Psi(x)$ near the point $x_j$. Then the transformed functions
\beq
\fj_{a;k}(y) := (U_a \fj_k)(y)
\eeq
will be equally optimal for interpolating $\Psi_a(y)$ near the point $y_j=\La\inv1(x_j)$. 

Assume that the interpolation hypothesis $\fj_k(x)$ of the original problem are
polynomials, i.e. the standard finite difference scheme of a given order. 
Then in almost all cases the equally optimal interpolation hypothesis 
$\fj_{a,k}(y)=\la(y)\inv{1/2}\fj_k(\La(y))$ for the transformed solution $\Psi_a(y)$
will {\em not} be polynomials and one needs to re-derive the corresponding finite difference scheme by 
Eq.~(\ref{eq:fdstandard}). 

In practice, the procedure for obtaining schemes for any transformed linear operator $B_a=U_aBU_a^*$
is very simple. Analogous to Eqs.~(\ref{eq:hyp0}) and (\ref{eq:fdstandard}) 
one writes
\beq\label{eq:hypa}
\Psi_a(y)=(U_a\Psi)(y)\approx \sum_{k=-P}^{P} (U_a\fj_k)(y) c_k=\sum_{k=-P}^{P} \la(y)^{1/2}(\fj_k(\La(y)) c_k
\eeq
\bea
(B_a \Psi_a)(y_j)&=&(U_a B U_a^*)\Psi_a(y_j)\approx \sum_{k=-P}^{P} (U_aB\fj_k)(y) c_k\nonumber\\
&=& 
\sum_{k=-P}^{P} \la(y_j)^{1/2}(B\fj_k)(\La(y_j)) c_k.
\label{eq:dera}
\eea
With the notation $\left(\mF\up{j}_a\right)_{lk}=\la(y_l)^{1/2}(\fj_k(\La(y_l))$
one obtains the adjusted scheme at point $y_j$ as
\beq\label{eq:transScheme}
b_{a;l}\up{j}=\sum_{k=1}^Q \la(y_j)^{1/2}(B\fj_k)(\La(y_j)) \left[\left(\mF\up{j}_a\right)\inv1\right]_{kl}.
\eeq 
For the first derivative $[\ddy]_a$ we insert $B\fj_k=\left[\fj_k\right]'$ into Eq.~(\ref{eq:transScheme}), 
for the second derivative   $[\ddyy]_a$ we insert $B\fj_k=\left[\fj_k\right]''$ etc.
As long as all transformations $U_a$ are unitary, one obtains symmetric schemes for $\vPsi_a$, provided
that the original problem gives symmetric schemes. 

The adjustment of $\fj_{a,k}$ is particularly important, when the transformation $\La$ is not analytical. This is
for example the case, when we want to switch from constant spacing $\Delta x_1$ in one region to a different 
constant spacing $\Delta x_2$ in some other region, possibly with a smooth, say linear, transition
in between. If we attempt to approximate the transformed solution $\Psi_a(y)$ by higher order polynomials
in $y$, i.e. when we use standard higher order FD schemes, we will observe a loss of the convergence order. 
Assuming the original $\Psi(x)$ has a convergent Taylor expansion, the solution \wrt to the 
transformed coordinate $\la^{1/2}(y)\Psi_a(y)=\Psi(\La(y))$ 
only has a continuous first derivative $\ddy$. Already the second derivative will be discontinuous and 
all higher derivatives show $\delta$-function like singularities
at the boundaries between constant and linearly changing spacing, which disqualifies any attempt of expanding 
into a convergent Taylor series. Making the transition smoother only postpones
the problem to higher orders. On the other hand, we will demonstrate below that adjusting $\fj_{a,k}$ and using Eq.~(\ref{eq:transScheme})
for the operators preserves the approximation order and no extra computational cost ensues 
apart from the initial construction of the local schemes $b_{a;l}\up{j}$. 

From the discussion it is clear that we can also use transformations generated by non-differentiable $\la(x)$, 
as they arise in ECS. All we need to do is to replace polynomials of standard FD schemes with 
their transformed counterparts. Finally, also the idea underlying the irECS discretization discussed above
can be transferred to FD schemes. Considering the approximately exponentially spaced Radau quadrature points appearing
in the FEM-DVR implementation of irECS, 
we see that the essential point of irECS is that the function is sampled at rapidly increasing spacing rather 
than uniformly. By uniting complex scaling with non-equidistant sampling we will obtain nearly as efficient 
absorption scheme with FD as irECS with FEM-DVR discretization.

\subsection{Absorption with equidistant FD grids}

We first investigate complex scaling with equidistant grids for the model Hamiltonian (\ref{eq:modelH}). 
We will show that comparable accuracies are reached for FEM-DVR and FD at equal grid sizes and equal orders $Q_n$.

\begin{figure}
\includegraphics[width=0.95\textwidth]{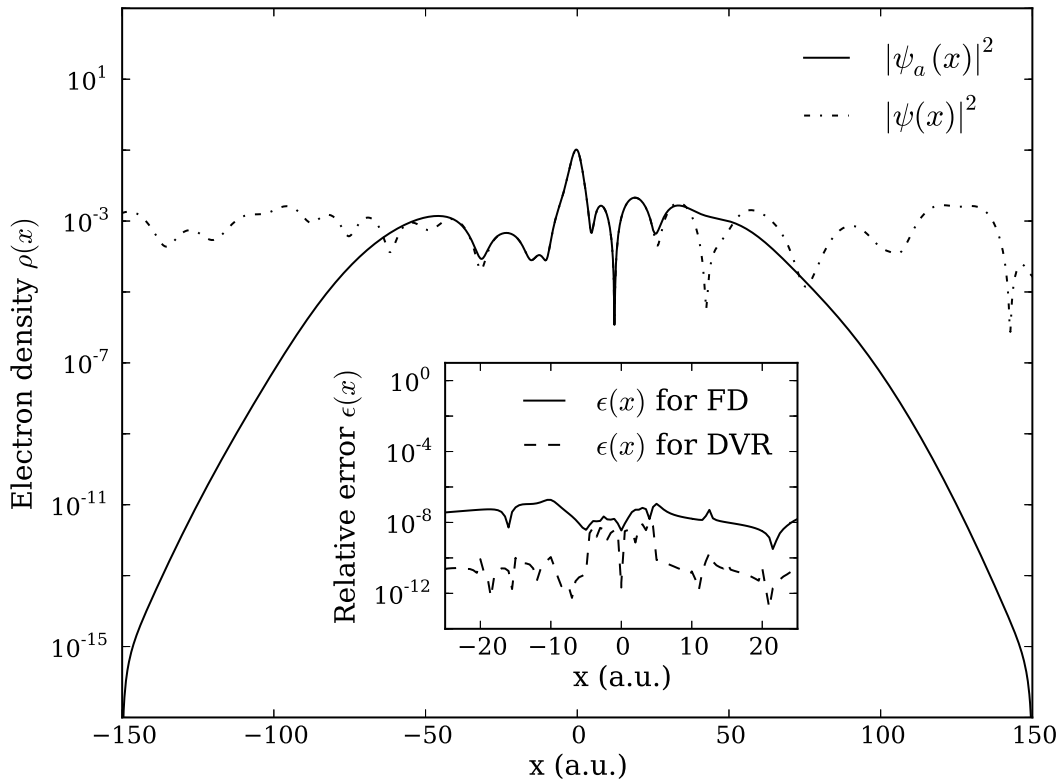}
\caption{\label{fig:dvrfd} 
Densities of the unscaled $|\Psi(x)|^2$ (dot-dashed line) and the complex scaled  $|\Psi_a(x)|^2$ system (solid line). Damping
by complex scaling at large $|x|$ is clearly exposed. Inset: relative errors $\ep(x)$ in the inner domain for DVR and FD at equal 
grid sizes and order.}
\end{figure}

We again use the transformation defined by Eq.~(\ref{eq:lambda}) with $a=\exp(i/2)$ and $R_0=25$ and 
FD schemes are constructed according to Eq.~(\ref{eq:transScheme}). The same constant order $Q_n\equiv21$ 
is used in FD and FEM-DVR calculations on 1200 grid points in a computational box $[-150,150]$.
Fig.~\ref{fig:dvrfd} compares complex scaled densities $|\Psi_a(x)|^2$ and errors of the two methods relative
to a converged calculation on a large box without using absorbing boundaries. Comparing unscaled $|\Psi(x)|^2$ with 
complex scaled $|\Psi_a(x)|$ densities one can clearly discern the suppression of the density to below $10^{-15}$ as the 
solution approaches the box boundaries. On the level of densities FD and FEM-DVR results are indistinguishable.
Also, relative to a fully converged unscaled calculation, errors in the inner region are comparable.

\begin{figure}
\includegraphics[width=0.45\textwidth]{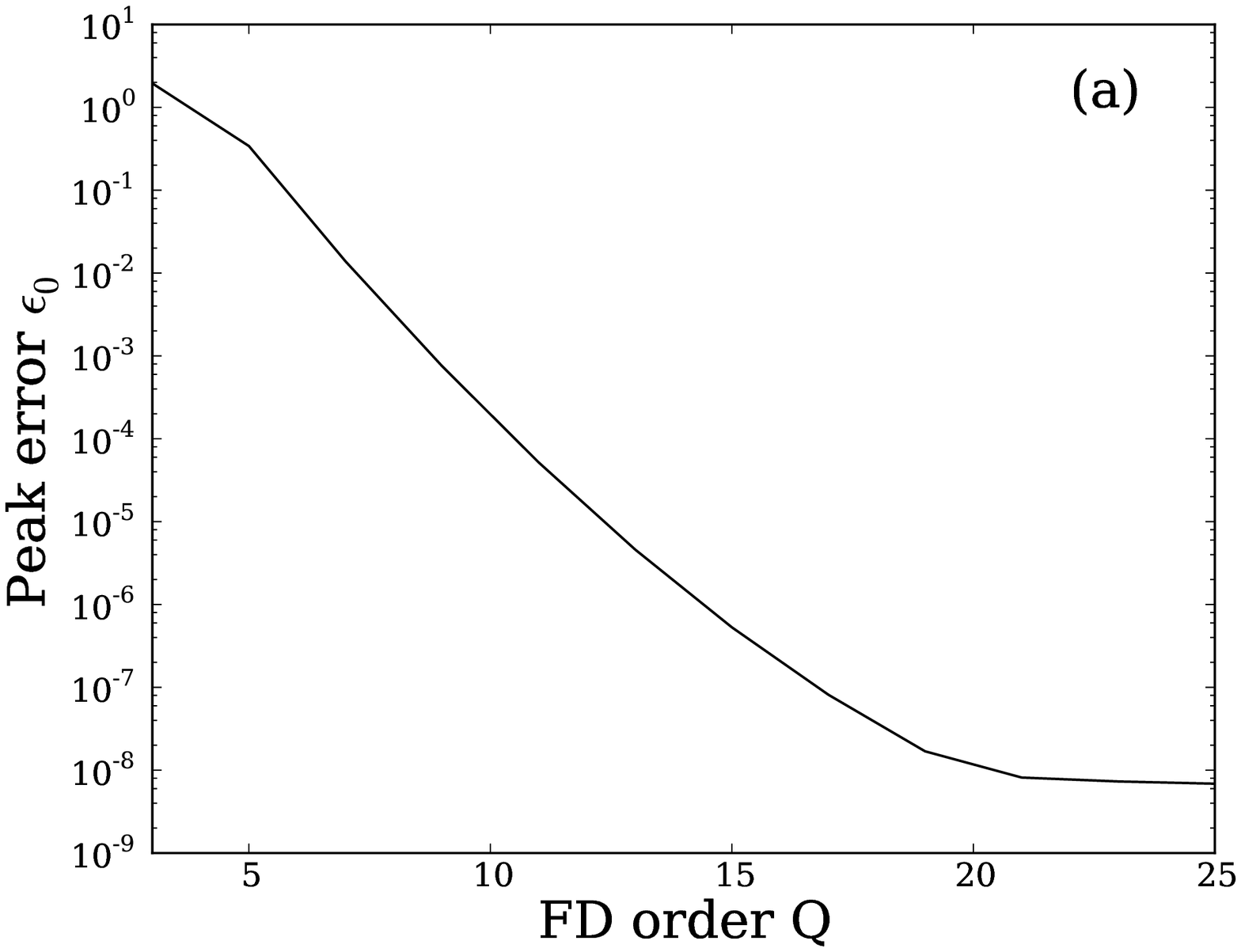}
\includegraphics[width=0.45\textwidth]{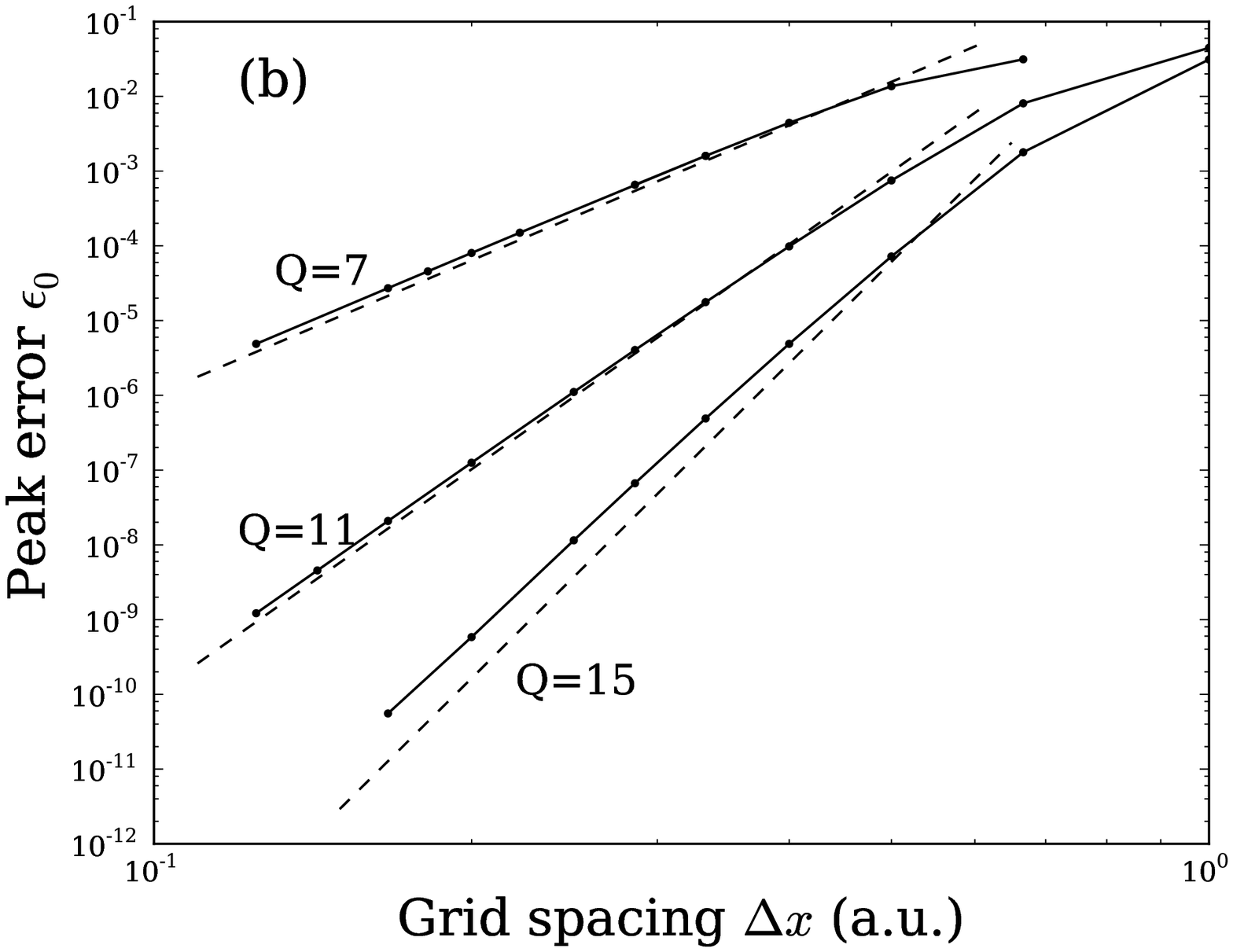}
\caption{\label{fig:fdorders} 
Convergence of peak relative error $\ep_0$ on the inner domain: (a) convergence with $Q$ for
1600 equidistant grid points on $[-200,200]$, (b) convergence with $\Delta x$, dashed lines $\Delta x^{Q-1}$.
At $Q=15$ the actual convergence is only $\approx \Delta x^{13}$. 
 }
\end{figure}

For completeness we show in Fig.~\ref{fig:fdorders} 
that the error drops exponentially with the FD order $Q$. This proves the full consistency
and that the non-differentiable nature of the complex scaled solution
is fully accounted for by the generalized FD scheme.

\subsection{Absorption with exponential FD grids}

Having established standard ECS for FD grids, as a last step we implement the idea underlying irECS also for 
FD schemes. As discussed, this consists in exponentially sampling the scaled solution $\Psi_a$ in the absorbing domain.  
In the notation introduced above, exponential sampling can be achieved by applying a unitary transformation 
to the complex scaled solution $\Psi_a(y)$:
\beq
\Psi_a(y)\to \Psi_{a,\ga}(z)= (W_\ga\Psi_a)(z),
\eeq
where $W_\ga$ is a real scaling transformation with 
\beq
\la(z)=\lcase
1 & \text{ for } |z|<R_0\\
\exp[\ga (|z|-R_0)] &\text{ for } |z|\geq R_0
\rcase.
\eeq

\begin{figure}
\includegraphics[width=0.95\textwidth]{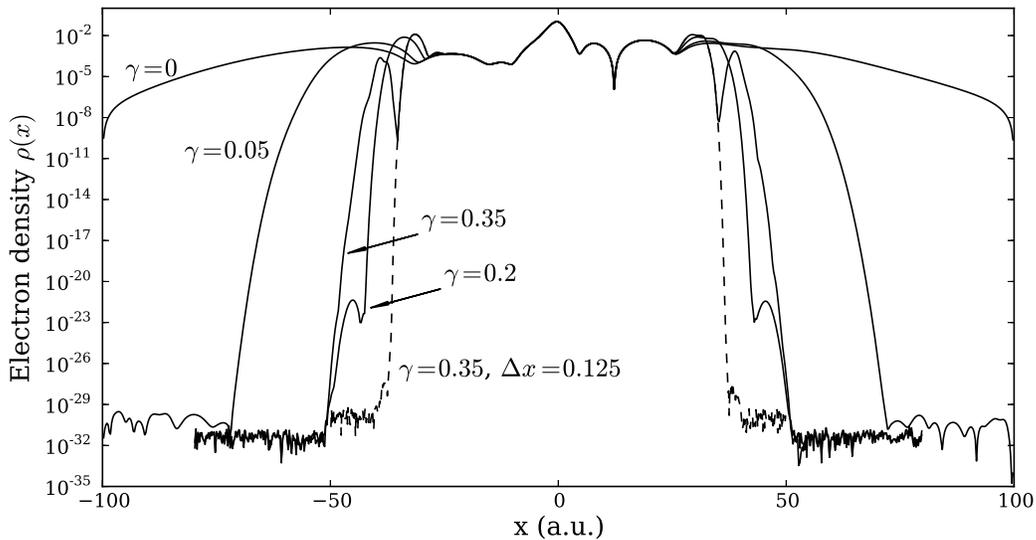}
\caption{\label{fig:fdirecs} 
The irECS method for FD schemes. Solid lines: equidistant grid $\ga=0$ and exponential scaling in the absorbing 
region.  Spacing of the transformed grid $\Delta z=0.25$ and order $Q=21$. 
At $\ga=0.35$ insufficient sampling leads to artefacts. Dashed line: with spacing 
$\Delta z=0.125$ exponent $\ga=0.125$ does not show artefacts.}
\end{figure}

For the comparisons we use a fixed complex scaling radius $R_0=25$ and fixed complex scaling phase $a=\exp(i/2)$ as above.
All complex scaled calculations are performed on the interval $[-100,100]$ with a total of 400 grid points.
This allows accuracies of the density of $\lesssim 10\inv7$ in unscaled domain $[-25,25]$, if absorption 
is perfect. Fig.~\ref{fig:fdirecs} shows
the complex scaled $|\Psi_a(y)|^2$ with equidistant grid and the complex scaled $|\Psi_{\ga,a}(z)|$ with exponential grids in the scaled domain
for $\ga=0.1,0.2,0.35$. The spatial damping of the solution
by the complex scaling transformation is clearly visible. Near the boundaries the equidistant grid solution $|\Psi_a|^2$ 
drops to $\sim 10\inv{9}$.
The effect of the Dirichlet boundary condition at $|y|=100$ is clearly visible, but has no consequences on the accuracy level 
of interest. 
The exponential grid maps the absorbing domain into smaller boundary layers outside $[-25,25]$. An optimum is reached near
$\ga=0.2$: the density drops to below the level $\lesssim 10\inv9$ for all $|z|>40$. Larger contraction to $\ga=0.35$ does not
lead to 
further gains: although the solution shrinks to smaller $|z|$ initially, the exponential spacing $x_j=x(z_j)$ 
is becoming too wide to represent the solution 
in the asymptotic region. An artefact on the level $\sim 10\inv5$ appears which causes reflection errors. 
As shown in the figure, the artefact can be fully suppressed by reducing 
the grid spacing, but this defeats the purpose of minimizing the number of points used for absorption.

\begin{figure}
\includegraphics[width=0.48\textwidth]{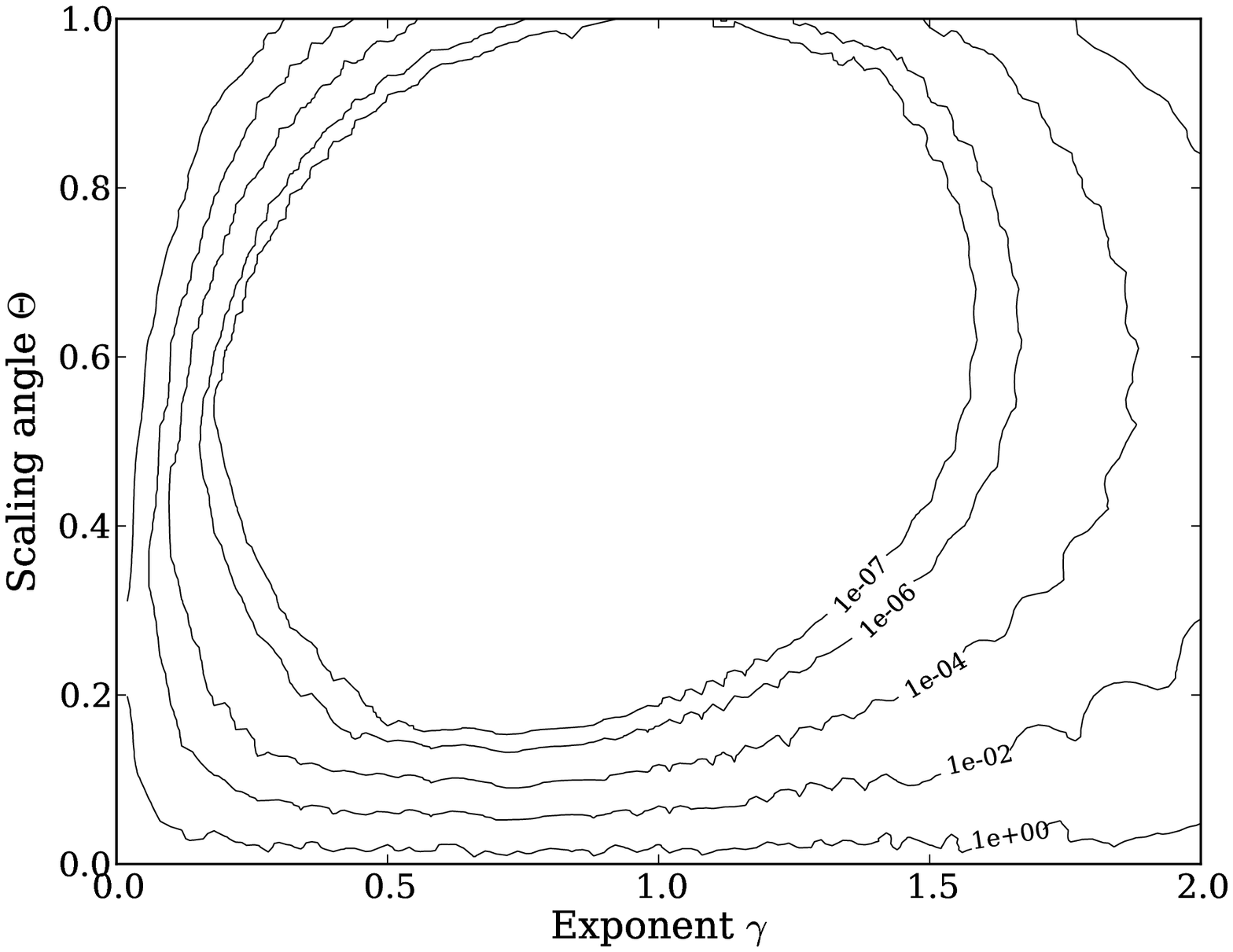}
\includegraphics[width=0.48\textwidth]{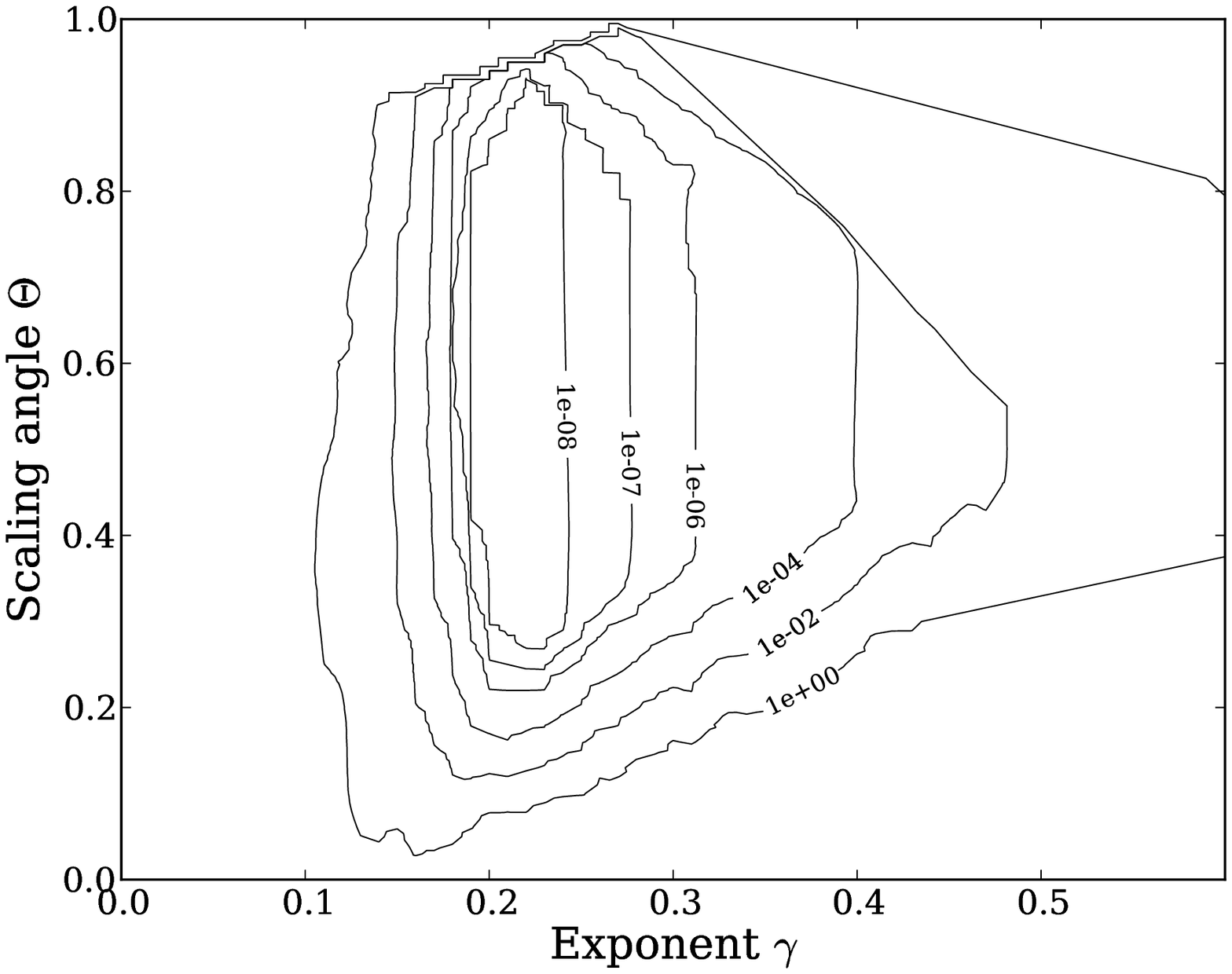}
\caption{\label{fig:gammatheta}
Relative error of FEM-DVR (left) and FD (right) schemes as a function of exponent $\ga$ and scaling angle  $\th$
at equal number of discretization points and comparable matrix sparsity.
The admissible $\ga$-range is narrower for FD, $\th$ can be chosen in a wide range with either method.
 }
\end{figure}

Finally, we investigate the dependence of absorption on $\ga$ and $\th$ for 
FD and FEM-DVR. In both methods we use the same number of 201 discretization
points in the unscaled domain $[-25,25]$ and the same order $Q=21$, i.e. a 21 point scheme for FD and degree 20 
polynomials for FEM-DVR, and we use the same number of $A=60$ points for absorption. At the grid spacing 
of $\Delta z=0.25$ this amount to Dirichlet boundary conditions at $|z|=40$ for FD. 
Note that this results in matrices the same size and of comparable sparsity in the inner region, with band-width 21 for FD and near block-diagonal 
matrices of blocks $21\times 21$ for FEM-DVR. Fig.~\ref{fig:gammatheta} shows the relative errors of $|\Psi(x)|^2$ in the inner 
domain as obtained with either method. In FD the range of admissible $\ga$ remains smaller than in FEM-DVR: as large $\gamma$ 
lead to a stronger contraction of the wave function, we see that the profit from the irECS idea is somewhat lower in FD.

\subsection{Smooth exterior scaling}
\begin{figure}
\includegraphics[width=0.95\textwidth]{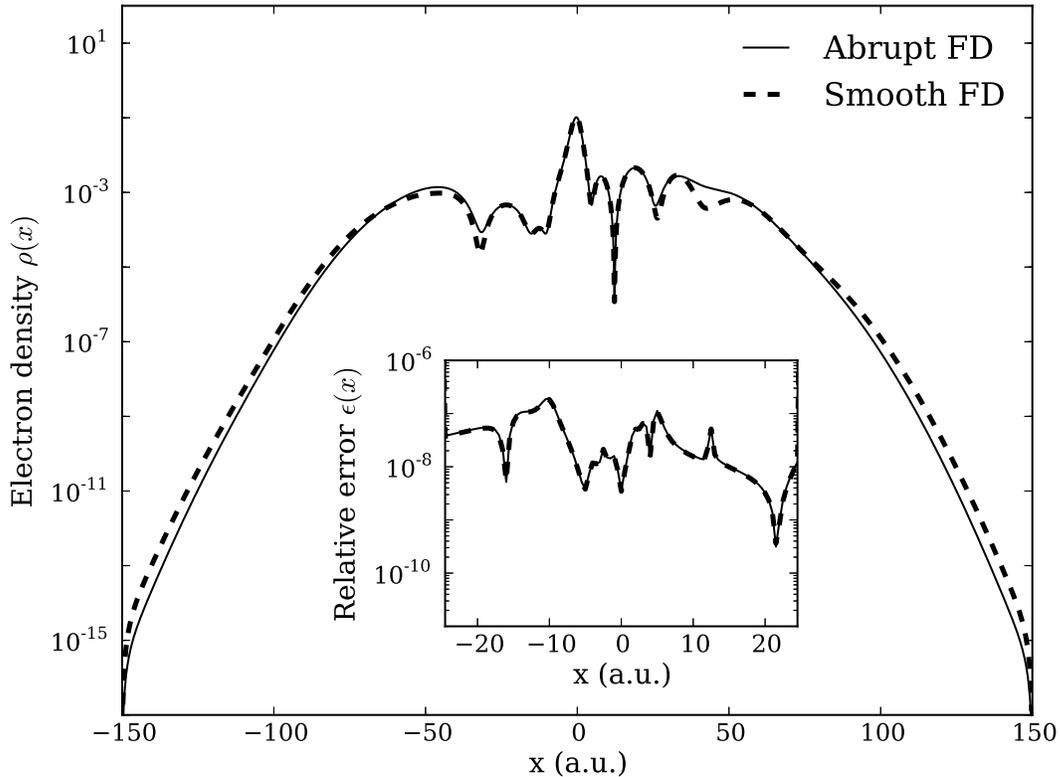}
\caption{\label{fig:smooth}
Abrupt (solid line) vs.\ smooth ECS (dashed line). Exponential suppression is slightly less with smooth scaling, errors in the unscaled 
region are identical.}
\end{figure}

We have demonstrated that the generalized FD schemes as well as FEM-DVR grids allow for discontinuous scaling functions $\la(x)$.
{\it A fortiori} we expect smooth exterior scaling to work correctly, if implemented by the above principles. This is indeed the
case. We use the scaling function
\beq
\la(y)=\lcase 
1 & \text{ for } |y|<R_0\\
1+(a-1)s(|y|) & \text{ for } R_0<|y|<R_1\\
a & \text{ for } |y|>R_1,
\rcase
\eeq 
where the $3^{rd}$ order polynomial $s(y)$ smoothly connects the inner domain with the region $|y|>R_1$ such that
$\la(y)$ is differentiable at $y=R_0$ and $y=R_1$.
Applying the same procedure as above and using $R_1=45$, i.e. smoothing over a range of $20\,au$, we find essentially
identical results as for the abrupt transition used earlier, see Fig.~\ref{fig:smooth}.

The usual motivation for smooth scaling is not to use the generalized schemes but to apply standard schemes. There is no
unique way for defining such a scheme for a polynomial interpolation hypothesis. There are two different
possibilities: one by bringing the scaled second derivative Eq.~(\ref{eq:ddxxTrans}) to a form that allows the use
of standard FD schemes for $\ddy$ and $\ddyy$ without increase of band-width
\beq\label{eq:explicitU}
[\ddxx]_a=U_a\ddxx U_a^*=\frac{1}{\la}\ddyy\frac{1}{\la}
-\frac12\frac{\la'}{\la^2}\ddy\frac{1}{\la}
+\frac{1}{\la}\ddy\frac12\frac{\la'}{\la^2}
-\frac14\left(\frac{\la'}{\la^2}\right)^2.
\eeq 
In a second approach, one can use the procedure for constructing finite difference schemes 
 that lead to Eq.~(\ref{eq:fdstandard}) for the complete operator (\ref{eq:explicitU}) with 
polynomial interpolation in place of the correctly scaled polynomials. 
The latter fully uses the polynomial interpolation hypothesis, the former introduces some additional approximations in
evaluating the transformed derivatives of the polynomials $[\ddxx]_ap_k(y)$. Note that in both 
cases the polynomial interpolation hypothesis is unjustified and the approximation cannot be successful
beyond the lowest orders.

\begin{figure}
\includegraphics[width=0.95\textwidth]{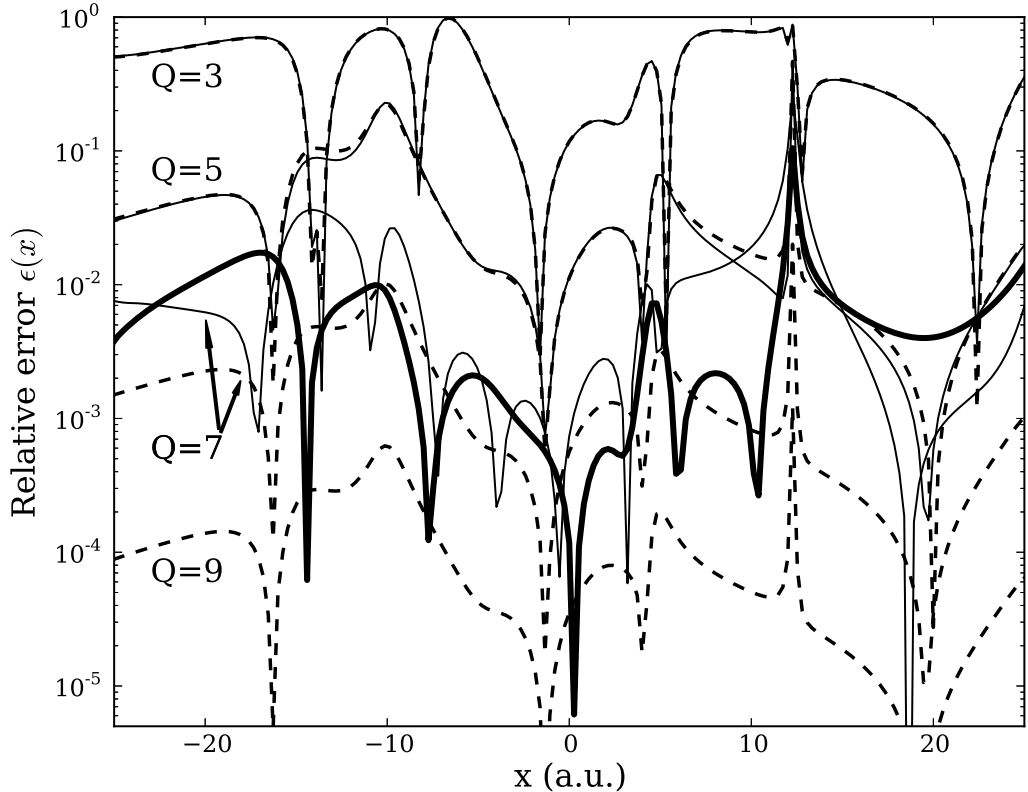}
\caption{\label{fig:standard}
Convergence with order $Q$ for scaled and standard  polynomial schemes. 
Smooth exterior scaling with a third degree polynomial smoothing function on an equidistant grid were used. 
Relative errors $\ep(x)$ in the unscaled
region $[-25,25]$ are show for $Q=3,5,7,9$. Scaled schemes errors decrease exponentially with $Q$ (dashed lines). 
For polynomial schemes (solid lines) no gain is made for $Q>7$ ). Calculations
are with $600$ points on the interval $[-80,80]$. The thick solid line is for a standard polynomial scheme
with $Q=7$ and as many as 4800 points on $[-80,80]$.}
\end{figure}

For the present demonstration we only study the second possibility, i.e.\ using the full, but incorrect
polynomial interpolation without making further approximations.
The limitation of convergence order for the polynomial interpolation hypothesis applied to a smoothly exterior
scaled grid
is illustrated in Fig.~\ref{fig:standard}. For orders Q=3 and Q=5 polynomial and scaled interpolation 
give essentially the same result. However, while the correctly scaled interpolations shows steady exponential convergence
at $Q=7,9$, no further gains can be made by increasing the order with the unscaled polynomial hypothesis. 
For higher accuracies one is forced to increase the of number grid points, which causes exceeding numerical
cost not only in terms of the problem size but also in stiffness of the equations.

Without demonstration, we only remark that also for the FEM-DVR scheme smooth transition does not bear any advantage 
over the abrupt transition used in the FEM-DVR calculations above.
On the contrary, the rather complicated smoothly scaled operators must be programmed and the finite modulation of
the solution in the transition region usually requires more grid points there, which in turn 
may raise the stiffness of the dynamical equations. 

\section{Conclusions}

With the present studies we have demonstrated that grid methods allow for highly efficient
absorption schemes. The irECS method had been originally developed using a finite element implementation, 
where also the far superior performance of irECS compared to multiplicative CAPs and MFA had been highlighted.
With the present extension of the method to FEM-DVR and FD grids one
has discretizations that are easy to implement, have low floating counts, and 
are straight forward for parallelization. The latter point may be the most important advantage.
While in transferring irECS to FEM-DVR no practical problems of any kind were encountered,
we had to derive a new approach to FD for complex scaled problems. In fact, the computation 
of scaled schemes is technically simple and provides schemes that are as nearly as efficient in terms
of discretization size as the FEM and FEM-DVR schemes of the same order. In particular, they can be pushed
to high orders with exponential reduction of errors to near machine precision.

We have shown that irECS can be considered as ECS with an exponential grid in 
the absorbing domain. Both, in FEM-DVR and in FD the transition from the equidistant unscaled to
the exponentially spaced absorbing region is numerically seamless, i.e. 
it does not produce any artefacts compared to an equidistant grid. The numerical gain by the
reduction of grid points is substantial and no increase in stiffness was observed.

We have further studied smooth exterior scaling and shown that one can smoothen the transition between scaled 
and unscaled region. This case was only discussed for the FD implementation, as one usually 
resorts to such a procedure to circumvent manifest problem that arises for using standard
FD schemes across abrupt transitions. Absorption
works for smooth scaling as well as it does for abrupt scaling, but only if scaled, non-polynomial 
schemes are used. With standard polynomial schemes, lack of differentiability leads to a strict limitation of
the consistency to the point where the corresponding higher derivatives of the smoothing 
function become singular. Smoothing does not bear any mathematical or computational advantage
as compared to abrupt transitions, on the contrary, it tends to complicate implementation
as the essentially arbitrary smoothing transformation must be incorporated into the scheme.
Therefore, at least from the present work, we would advice to use abrupt changeovers
wherever possible.

The FD schemes for non-equidistant grids, in particular the abrupt and reflectionless 
transition between grid spacings occurring in abrupt exterior scaling, 
may be of broad interest for uses beyond the problem of absorbing boundaries considered here.
Locally adapted FD  grids are frequently used in literature. To the best of our knowledge, the approach presented
here and proven to fully maintain convergence orders is novel.
In future work, we plan to investigate the problem for non-equidistant
grid representations of Maxwell's equations.

\section*{Acknowledgment}
We acknowledge support by the excellence cluster ``Munich
Center for Advanced Photonics (MAP)'' and by the Austrian Science
Foundation project ViCoM (F41).

\bibliography{/home/scrinzi/Papers/bibliography/photonics_theory}{}

\begin{thebibliography}{21}
\expandafter\ifx\csname natexlab\endcsname\relax\def\natexlab#1{#1}\fi
\expandafter\ifx\csname bibnamefont\endcsname\relax
  \def\bibnamefont#1{#1}\fi
\expandafter\ifx\csname bibfnamefont\endcsname\relax
  \def\bibfnamefont#1{#1}\fi
\expandafter\ifx\csname citenamefont\endcsname\relax
  \def\citenamefont#1{#1}\fi
\expandafter\ifx\csname url\endcsname\relax
  \def\url#1{\texttt{#1}}\fi
\expandafter\ifx\csname urlprefix\endcsname\relax\def\urlprefix{URL }\fi
\providecommand{\bibinfo}[2]{#2}
\providecommand{\eprint}[2][]{\url{#2}}

\bibitem[{\citenamefont{McCurdy et~al.}(1991)\citenamefont{McCurdy, Stroud, and
  Wisinski}}]{mccurdy91:complex_scaling_tdse}
\bibinfo{author}{\bibfnamefont{C.~W.} \bibnamefont{McCurdy}},
  \bibinfo{author}{\bibfnamefont{C.~K.} \bibnamefont{Stroud}},
  \bibnamefont{and} \bibinfo{author}{\bibfnamefont{M.~K.}
  \bibnamefont{Wisinski}}, \bibinfo{journal}{Phys. Rev. A}
  \textbf{\bibinfo{volume}{43}}, \bibinfo{pages}{5980} (\bibinfo{year}{1991}).

\bibitem[{\citenamefont{Scrinzi}(2010)}]{scrinzi10:irecs}
\bibinfo{author}{\bibfnamefont{A.}~\bibnamefont{Scrinzi}},
  \bibinfo{journal}{Phys. Rev. A} \textbf{\bibinfo{volume}{81}},
  \bibinfo{pages}{053845} (\bibinfo{year}{2010}).

\bibitem[{\citenamefont{Scrinzi et~al.}(2014)\citenamefont{Scrinzi, Stimming,
  and Mauser}}]{scrinzi14:ecs-pml}
\bibinfo{author}{\bibfnamefont{A.}~\bibnamefont{Scrinzi}},
  \bibinfo{author}{\bibfnamefont{H.~P.} \bibnamefont{Stimming}},
  \bibnamefont{and} \bibinfo{author}{\bibfnamefont{N.~J.}
  \bibnamefont{Mauser}}, \bibinfo{journal}{J. Comput. Physics} pp.
  \bibinfo{pages}{98--107} (\bibinfo{year}{2014}).

\bibitem[{\citenamefont{Telnov et~al.}(2013)\citenamefont{Telnov, Sosnova,
  Rozenbaum, and Chu}}]{telnov13:ecs}
\bibinfo{author}{\bibfnamefont{D.~A.} \bibnamefont{Telnov}},
  \bibinfo{author}{\bibfnamefont{K.~E.} \bibnamefont{Sosnova}},
  \bibinfo{author}{\bibfnamefont{E.}~\bibnamefont{Rozenbaum}},
  \bibnamefont{and} \bibinfo{author}{\bibfnamefont{S.-I.} \bibnamefont{Chu}},
  \bibinfo{journal}{Phys. Rev. A} \textbf{\bibinfo{volume}{87}},
  \bibinfo{pages}{053406} (\bibinfo{year}{2013}),
  \urlprefix\url{http://link.aps.org/doi/10.1103/PhysRevA.87.053406}.

\bibitem[{\citenamefont{Dujardin et~al.}(2014)\citenamefont{Dujardin, Saenz,
  and Schlagheck}}]{dujardin14:ecs}
\bibinfo{author}{\bibfnamefont{J.}~\bibnamefont{Dujardin}},
  \bibinfo{author}{\bibfnamefont{A.}~\bibnamefont{Saenz}}, \bibnamefont{and}
  \bibinfo{author}{\bibfnamefont{P.}~\bibnamefont{Schlagheck}},
  \bibinfo{journal}{Applied Physics B} \textbf{\bibinfo{volume}{117}},
  \bibinfo{pages}{765} (\bibinfo{year}{2014}), ISSN \bibinfo{issn}{0946-2171},
  \urlprefix\url{http://dx.doi.org/10.1007/s00340-014-5804-3}.

\bibitem[{\citenamefont{De~Giovannini et~al.}(2015)\citenamefont{De~Giovannini,
  Larsen, and Rubio}}]{deGiovanni15:absorbers}
\bibinfo{author}{\bibfnamefont{U.}~\bibnamefont{De~Giovannini}},
  \bibinfo{author}{\bibfnamefont{A.}~\bibnamefont{Larsen}}, \bibnamefont{and}
  \bibinfo{author}{\bibfnamefont{A.}~\bibnamefont{Rubio}},
  \bibinfo{journal}{The European Physical Journal B}
  \textbf{\bibinfo{volume}{88}}, \bibinfo{eid}{56} (\bibinfo{year}{2015}), ISSN
  \bibinfo{issn}{1434-6028},
  \urlprefix\url{http://dx.doi.org/10.1140/epjb/e2015-50808-0}.

\bibitem[{\citenamefont{Miller et~al.}(2014)\citenamefont{Miller,
  Hern\'andez-Garc\'{i}a, Jar\'{o}n-Becker, and Becker}}]{miller14:ecs}
\bibinfo{author}{\bibfnamefont{M.~R.} \bibnamefont{Miller}},
  \bibinfo{author}{\bibfnamefont{C.}~\bibnamefont{Hern\'andez-Garc\'{i}a}},
  \bibinfo{author}{\bibfnamefont{A.}~\bibnamefont{Jar\'{o}n-Becker}},
  \bibnamefont{and} \bibinfo{author}{\bibfnamefont{A.}~\bibnamefont{Becker}},
  \bibinfo{journal}{Phys. Rev. A} \textbf{\bibinfo{volume}{90}},
  \bibinfo{pages}{053409} (\bibinfo{year}{2014}),
  \urlprefix\url{http://link.aps.org/doi/10.1103/PhysRevA.90.053409}.

\bibitem[{\citenamefont{Saenz}()}]{saenz:private}
\bibinfo{author}{\bibfnamefont{A.}~\bibnamefont{Saenz}}, \bibinfo{note}{private
  communication}.

\bibitem[{\citenamefont{Rom et~al.}(1990)\citenamefont{Rom, Engdahl, and
  Moiseyev}}]{rom90:smoothECS}
\bibinfo{author}{\bibfnamefont{N.}~\bibnamefont{Rom}},
  \bibinfo{author}{\bibfnamefont{E.}~\bibnamefont{Engdahl}}, \bibnamefont{and}
  \bibinfo{author}{\bibfnamefont{N.}~\bibnamefont{Moiseyev}},
  \bibinfo{journal}{J. Chem. Phys.} \textbf{\bibinfo{volume}{93}},
  \bibinfo{pages}{3413} (\bibinfo{year}{1990}).

\bibitem[{tRe()}]{tRecXweb}
\emph{\bibinfo{title}{The {tRecX} {Homepage}}},
  \urlprefix\url{http://homepages.physik.uni-muenchen.de/~armin.scrinzi/tRecX}.

\bibitem[{\citenamefont{Reed and Simon}(1982)}]{reed_simon82:complex_scaling}
\bibinfo{author}{\bibfnamefont{M.}~\bibnamefont{Reed}} \bibnamefont{and}
  \bibinfo{author}{\bibfnamefont{B.}~\bibnamefont{Simon}},
  \emph{\bibinfo{title}{Methods of Modern Mathematical Physics}}
  (\bibinfo{publisher}{Academic, New York}, \bibinfo{year}{1982}), p.
  \bibinfo{pages}{183{ff.}}

\bibitem[{\citenamefont{Manolopoulos and Wyatt}(1988)}]{manolopoulos88:femdvr}
\bibinfo{author}{\bibfnamefont{D.}~\bibnamefont{Manolopoulos}}
  \bibnamefont{and} \bibinfo{author}{\bibfnamefont{R.}~\bibnamefont{Wyatt}},
  \bibinfo{journal}{Chemical Physics Letters} \textbf{\bibinfo{volume}{152}},
  \bibinfo{pages}{23 } (\bibinfo{year}{1988}), ISSN \bibinfo{issn}{0009-2614},
  \urlprefix\url{http://www.sciencedirect.com/science/article/pii/0009261488873226}.

\bibitem[{\citenamefont{Scrinzi and Elander}(1993)}]{scrinzi:jcp1993}
\bibinfo{author}{\bibfnamefont{A.}~\bibnamefont{Scrinzi}} \bibnamefont{and}
  \bibinfo{author}{\bibfnamefont{N.}~\bibnamefont{Elander}},
  \bibinfo{journal}{J. Chem. Phys.} \textbf{\bibinfo{volume}{98}},
  \bibinfo{pages}{3866} (\bibinfo{year}{1993}).

\bibitem[{\citenamefont{Hofmann et~al.}({2014})\citenamefont{Hofmann, Landsman,
  Zielinski, Cirelli, Zimmermann, Scrinzi, and Keller}}]{hofmann14:elliptic}
\bibinfo{author}{\bibfnamefont{C.}~\bibnamefont{Hofmann}},
  \bibinfo{author}{\bibfnamefont{A.~S.} \bibnamefont{Landsman}},
  \bibinfo{author}{\bibfnamefont{A.}~\bibnamefont{Zielinski}},
  \bibinfo{author}{\bibfnamefont{C.}~\bibnamefont{Cirelli}},
  \bibinfo{author}{\bibfnamefont{T.}~\bibnamefont{Zimmermann}},
  \bibinfo{author}{\bibfnamefont{A.}~\bibnamefont{Scrinzi}}, \bibnamefont{and}
  \bibinfo{author}{\bibfnamefont{U.}~\bibnamefont{Keller}},
  \bibinfo{journal}{{Phys. Rev. A}} \textbf{\bibinfo{volume}{{90}}}
  (\bibinfo{year}{{2014}}), ISSN \bibinfo{issn}{{1050-2947}}.

\bibitem[{\citenamefont{Majety et~al.}({2015})\citenamefont{Majety, Zielinski,
  and Scrinzi}}]{majety15:hacc}
\bibinfo{author}{\bibfnamefont{V.~P.} \bibnamefont{Majety}},
  \bibinfo{author}{\bibfnamefont{A.}~\bibnamefont{Zielinski}},
  \bibnamefont{and} \bibinfo{author}{\bibfnamefont{A.}~\bibnamefont{Scrinzi}},
  \bibinfo{journal}{{New. J. Phys.}} \textbf{\bibinfo{volume}{{17}}}
  (\bibinfo{year}{{2015}}), ISSN \bibinfo{issn}{{1367-2630}}.

\bibitem[{\citenamefont{Majety and Scrinzi}(2015)}]{majety15:exchange}
\bibinfo{author}{\bibfnamefont{V.~P.} \bibnamefont{Majety}} \bibnamefont{and}
  \bibinfo{author}{\bibfnamefont{A.}~\bibnamefont{Scrinzi}},
  \bibinfo{journal}{Phys. Rev. Lett.} \textbf{\bibinfo{volume}{115}},
  \bibinfo{pages}{103002} (\bibinfo{year}{2015}),
  \urlprefix\url{http://link.aps.org/doi/10.1103/PhysRevLett.115.103002}.

\bibitem[{\citenamefont{Zielinski et~al.}(2014)\citenamefont{Zielinski, Majety,
  Nagele, Pazourek, Burgd{\"o}rfer, and Scrinzi}}]{zielinski14:fanoArXiv}
\bibinfo{author}{\bibfnamefont{A.}~\bibnamefont{Zielinski}},
  \bibinfo{author}{\bibfnamefont{V.~P.} \bibnamefont{Majety}},
  \bibinfo{author}{\bibfnamefont{S.}~\bibnamefont{Nagele}},
  \bibinfo{author}{\bibfnamefont{R.}~\bibnamefont{Pazourek}},
  \bibinfo{author}{\bibfnamefont{J.}~\bibnamefont{Burgd{\"o}rfer}},
  \bibnamefont{and} \bibinfo{author}{\bibfnamefont{A.}~\bibnamefont{Scrinzi}},
  \bibinfo{journal}{arXiv:1405.4279 [physics.atom-ph]}  (\bibinfo{year}{2014}).

\bibitem[{\citenamefont{Torlina et~al.}(2015)\citenamefont{Torlina, Morales,
  Kaushal, Ivanov, Kheifets, Zielinski, Scrinzi, Muller, Sukiasyan, Ivanov
  et~al.}}]{torlina15:attoclock}
\bibinfo{author}{\bibfnamefont{L.}~\bibnamefont{Torlina}},
  \bibinfo{author}{\bibfnamefont{F.}~\bibnamefont{Morales}},
  \bibinfo{author}{\bibfnamefont{J.}~\bibnamefont{Kaushal}},
  \bibinfo{author}{\bibfnamefont{I.}~\bibnamefont{Ivanov}},
  \bibinfo{author}{\bibfnamefont{A.}~\bibnamefont{Kheifets}},
  \bibinfo{author}{\bibfnamefont{A.}~\bibnamefont{Zielinski}},
  \bibinfo{author}{\bibfnamefont{A.}~\bibnamefont{Scrinzi}},
  \bibinfo{author}{\bibfnamefont{H.}~\bibnamefont{Muller}},
  \bibinfo{author}{\bibfnamefont{S.}~\bibnamefont{Sukiasyan}},
  \bibinfo{author}{\bibfnamefont{M.}~\bibnamefont{Ivanov}},
  \bibnamefont{et~al.}, \bibinfo{journal}{Nature Physics}
  \textbf{\bibinfo{volume}{11}}, \bibinfo{pages}{503} (\bibinfo{year}{2015}).

\bibitem[{\citenamefont{Tao et~al.}(2009)\citenamefont{Tao, Vanroose, Reps,
  Rescigno, and McCurdy}}]{tao09:complex_scaling}
\bibinfo{author}{\bibfnamefont{L.}~\bibnamefont{Tao}},
  \bibinfo{author}{\bibfnamefont{W.}~\bibnamefont{Vanroose}},
  \bibinfo{author}{\bibfnamefont{B.}~\bibnamefont{Reps}},
  \bibinfo{author}{\bibfnamefont{T.~N.} \bibnamefont{Rescigno}},
  \bibnamefont{and} \bibinfo{author}{\bibfnamefont{C.~W.}
  \bibnamefont{McCurdy}}, \bibinfo{journal}{Phys. Rev. A}
  \textbf{\bibinfo{volume}{80}}, \bibinfo{pages}{063419}
  (\bibinfo{year}{2009}).

\bibitem[{\citenamefont{Zielinski et~al.}(2015)\citenamefont{Zielinski, Majety,
  and Scrinzi}}]{zielinski15:di}
\bibinfo{author}{\bibfnamefont{A.}~\bibnamefont{Zielinski}},
  \bibinfo{author}{\bibfnamefont{V.~P.} \bibnamefont{Majety}},
  \bibnamefont{and} \bibinfo{author}{\bibfnamefont{A.}~\bibnamefont{Scrinzi}},
  \bibinfo{journal}{(in preparation)}  (\bibinfo{year}{2015}).

\bibitem[{\citenamefont{Combes et~al.}(1987)\citenamefont{Combes, Duclos,
  Klein, and Seiler}}]{combes87:resonance}
\bibinfo{author}{\bibfnamefont{J.}~\bibnamefont{Combes}},
  \bibinfo{author}{\bibfnamefont{P.}~\bibnamefont{Duclos}},
  \bibinfo{author}{\bibfnamefont{M.}~\bibnamefont{Klein}}, \bibnamefont{and}
  \bibinfo{author}{\bibfnamefont{R.}~\bibnamefont{Seiler}},
  \bibinfo{journal}{Communications in Mathematical Physics}
  \textbf{\bibinfo{volume}{110}}, \bibinfo{pages}{215} (\bibinfo{year}{1987}),
  ISSN \bibinfo{issn}{0010-3616},
  \urlprefix\url{http://dx.doi.org/10.1007/BF01207364}.

\end{thebibliography}

\end{document}